\documentclass[12pt]{iopart}

 \expandafter\let\csname equation*\endcsname\relax
 \expandafter\let\csname endequation*\endcsname\relax
 \usepackage{amsmath}
 \usepackage{graphicx}   
 \usepackage{bbold}
 \usepackage{cite} 
 \usepackage{color}      
 \usepackage{ulem}

\newcommand{\braket}[2]{\left<#1|#2\right>}
\newcommand{\ket}[1]{| #1 \rangle}
\newcommand{\bra}[1]{\langle #1 |}

\begin{document}

\title[From quantum heat engines to laser cooling]{From quantum heat engines to laser cooling: Floquet theory beyond the Born-Markov approximation}
\author{Sebastian Restrepo$^1$, Javier Cerrillo$^1$, Philipp Strasberg$^2$ and Gernot Schaller$^1$}
\address{$^1$ Institut f\"ur Theoretische Physik, Technische Universit\"at Berlin, Hardenbergstr. 36, 10623 Berlin, Germany}
\address{$^2$ Complex Systems and Statistical Mechanics, Physics and Materials Science, University of Luxembourg, L-1511 Luxembourg, Luxembourg}
\ead{s.restrepo@tu-berlin.de}

\vspace{10pt}
\begin{indented}
\item[]December 2017
\end{indented}

\begin{abstract}
We combine the formalisms of Floquet theory and full counting statistics with a Markovian embedding strategy to access the
dynamics and thermodynamics of a periodically driven thermal machine beyond the conventional Born-Markov approximation. 
The working medium is a two-level system and we drive the tunneling as well as the coupling to one bath with the same 
period. We identify four different operating regimes of our machine which include a heat engine and a refrigerator. As the coupling strength with one bath is increased, the refrigerator regime disappears, the heat engine regime narrows and their efficiency and coefficient of performance decrease. Furthermore, our model can reproduce the setup of laser cooling of trapped ions in a specific parameter limit.
\end{abstract}

%
%
%
%
%

\section{Introduction}
In the recent years, notable progress has been made towards the experimental realization of small-scale thermal machines 
\cite{Hugel2002, Pekola2015, Martinez2016, Krishnamurthy2016, Roßnagel2016}. Specific examples in the quantum regime include a quantum absorption refrigerator with trapped ions \cite{Maslennikov2017} and quantum heat engines using an ensemble of nitrogen-vacancy centres \cite{Klatzow2017}. 
The theoretical study of these quantum thermal machines has in general been restricted to the weak coupling and Markovian regime \cite{Spohn1978a, Alicki1979, Kosloff2014}. This approach follows the logic of traditional thermodynamics, which is a weak coupling theory and relies on the distinction of systems interacting with each other due to the negligible contribution of their interface compared to the bulk properties \cite{Callen1985}. Nevertheless, this assumption becomes increasingly questionable at the nanoscale where, as the volume of systems becomes very small, boundaries cannot be clearly distinguished. In the present work we will present a formalism which allows to formulate a consistent thermodynamic framework for periodically driven systems coupled to multiple heat reservoirs -- even when the coupling is strong, driven and induces non-Markovian behaviour.
 
Time-periodic systems have been successfully studied by means of Floquet theory \cite{Shirley1965, Sambe1973, Grifoni1998} and a broad diversity of interesting phenomena such as coherent destruction of tunnelling \cite{Grossmann1991, Grossmann1992}, dynamical localisation \cite{Dunlap1986} and creation of new phases of matter \cite{Eckardt2005, Lindner2011, Bastidas2012} have been discovered. When connected to a heat reservoir, the steady-state dynamics of 
time-periodic quantum systems has been mainly described using the Floquet-Markov approach \cite{Kohler1997, Grifoni1998, Gelbwaser-Klimovsky2013, Kosloff2013, Szczygielski2013, Szczygielski2014, Gelbwaser-Klimovsky2015a}. 
It consists in deriving a weak coupling master equation \cite{Breuer2002} in the Floquet basis of the driven system. Under the secular approximation, conditions for the emergence of a Floquet-Gibbs state have been studied recently in \cite{Shirai2015}, a full stochastic thermodynamic analysis is given in \cite{Cuetara2015} and a thermodynamic analysis of laser cooling experiments by collisional redistribution \cite{Vogl2009, Vogl2011} can be found in \cite{Gelbwaser-Klimovsky2015a}.

To access the dynamics or study thermal transport in periodically driven systems beyond the Born-Markov approximation, more sophisticated methods must be used such as stochastic Liouville-von Neumann equations \cite{Schmidt2015}, perturbative high-frequency expansions \cite{Restrepo2016} or influence functional integral methods \cite{Carrega2016}. Their thermodynamic interpretation, however, is not always clear. Recent studies beyond the Born-Markov approximation with a consistent thermodynamic interpretation include non-equilibrium Green's functions \cite{Esposito2015, Bruch2016}, redefined system-reservoir partitions (collective coordinate mappings) \cite{Strasberg2016,Newman2017,Strasberg2017a,Schaller2017,Strasberg2017}, a quantum absorption refrigerator \cite{Mu2017}, quenched thermodynamic protocols \cite{Gallego2014,Perarnau-Llobet2017} and the derivation of an exact expression for the entropy production of a finite arbitrary system in contact with one or several thermal reservoirs \cite{Esposito2010}, but none of which were applied to periodically driven machines so far. Exceptions are \cite{Gelbwaser-Klimovsky2015} and \cite{Kato2016}. In \cite{Gelbwaser-Klimovsky2015} a polaron transformation and Floquet theory were used to include arbitrary strong coupling effects but only as long as the rotating wave approximation for the system Hamiltonian and the Markovian approximation for the reservoir hold. In \cite{Kato2016} a periodically driven three-level heat engine was studied using the numerical method of hierarchy of equations of motion, an approach that is based in a decomposition of the bath correlation function rather than an explicit representation of the bath. For this reason, access to more general, thermodynamically relevant thermal transport properties requires a tailored solution \cite{Cerrillo2016}.

However, none of the aforementioned methods was successfully applied to the combinations of problems we want to tackle: a driven system coupled to multiple heat reservoirs with a possibly strong, driven and non-Markovian coupling. To achieve this we combine three methods: a collective coordinate (CC) mapping \cite{Martinazzo2011,Woods2014}, Floquet theory for open systems \cite{Grifoni1998,Kohler1997} and full counting statistics \cite{Esposito2009}. This novel and unique combination provides access to steady state thermodynamics lifting the restriction of common assumptions such as a very fast driving, Markovian and weakly coupled reservoirs, and the secular approximation. First, in the CC mapping, we identify a collective degree of freedom in the reservoir, sometimes referred to as reaction coordinate \cite{Garg1985} that is responsible for the strong coupling and non-Markovian effects. Second, using Floquet theory, we derive a master equation for the original system plus CC and apply full counting statistics methods to obtain the change in energy of the reservoirs unambiguously. This strategy allows us to perform a consistent thermodynamic analysis of periodically driven thermal machines. Furthermore, it also allows us to accurately treat periodic time dependencies in the interaction between system and reservoirs, capturing their thermodynamic effects, which are inaccessible with standard methods. It is worth mentioning that even though the master equation that will be used to study the original system plus CC is Markovian, once the degrees of freedom of the CC are traced out, the dynamics of the system is non-Markovian.

This work is organized as follows: We start by briefly presenting the CC mapping (Sec.~\ref{sec:CC}), explaining the 
Floquet master equation used including the technique of counting fields (Sec.~\ref{sec:Floquet}) and defining important thermodynamic quantities (Sec.~\ref{sec:Thermo}).  In Sec.~\ref{sec:Results} we begin by specifying the  model for study.  
Results are then presented for the weak coupling and non-Markovian regime (Sec.~\ref{sec:weak}) with the identification 
of the refrigerator and heat engine regimes. In Sec. \ref{sec:strong} we explore the regime of finite coupling
 and establish the existence of two additional modes of operation, whereas in Sec.~\ref{sec:performance} we look at the performance  of our thermal machine. Finally, we study the limit in which we reproduce the setup of laser cooling (Sec.~\ref{sec:single_reservoir}) and follow with conclusions (Sec.~\ref{sec:Conclu}).

\section{Method}\label{sec:Method}
Our formalism is in general applicable to situations modeled by a system-bath Hamiltonian of the form
\begin{equation}\label{eq:H_xp}
H = H_S(t) + \sum_\nu \left[  H_{B}^{(\nu)} + H_{SB}^{(\nu)}(t) \right] , 
\end{equation}
where $H_S(t)$ is the Hamiltonian of the system, $H_B^{(\nu)}$ the Hamiltonian of reservoir $\nu$ and $H_{SB}^{(\nu)}(t)$ describes their interaction.
The heat reservoirs are described by a typical Brownian motion Hamiltonian as
\begin{equation}
H_{B}^{(\nu)} + H_{SB}^{(\nu)}(t) = \frac{1}{2}\sum\limits_k\left[p_{k\nu}^2+\omega_{k\nu}^2\left(x_{k\nu}-\frac{c_{k\nu}}{\omega_{k\nu}^2}S_{\nu}(t)\right)^2\right], 
\end{equation}
with frequency $\omega_{k\nu}$, mass-weighted positions $x_{k\nu}$ and momenta $p_{k\nu}$ of the bath oscillators and $S_{\nu}(t)$ a system operator which mediates the coupling to the bath with strength $c_{k\nu}$. We consider the case of a periodic time dependence in the working medium $H_S(t)=H_S(t+T)$ and coupling operator $S_{\nu}(t)=S_{\nu}(t+T)$, with period $T= 2 \pi / \omega_L$. The periodicity will allow us to study the problem by using Floquet theory.

To fully characterize the model, the spectral density of the heat reservoirs, defined by
\begin{equation}
J_{\nu}(\omega) \equiv \frac{\pi}{2}\sum\limits_k \frac{c_{k\nu}^2}{\omega_{k\nu}}\delta(\omega - \omega_{k\nu})  ,
\end{equation}
must be parametrized. The spectral density is a positive function for $\omega> 0$ and fulfills 
$J_{\nu}(\omega)\rightarrow 0$ for $\omega\rightarrow 0$ and $\omega \rightarrow \infty$. A structure-less spectral density (e.g. linear form) usually allows for a Markovian treatment of the reservoirs due to the fast 
decay of its associated correlation functions, while a more structured spectral density (e.g. strongly peaked around a 
frequency $\omega_{res}$) demands a more elaborate treatment. In the following, we will consider the case of a hot ($\nu=h$) and cold ($\nu=c$) reservoir where only the cold spectral density has a structured form:
\begin{equation}
J_{c}(\omega) = \frac{d_c^2 \gamma \omega}{(\omega^2-\omega_{res}^2)^2 + \gamma^2 \omega^2•}  , \label{eq:CSD}
 \qquad J_{h}(\omega) = \frac{d_h}{\omega_0} \omega  .
\end{equation}
\begin{figure}[t!]
\begin{center}
  \includegraphics[width=0.49\linewidth]{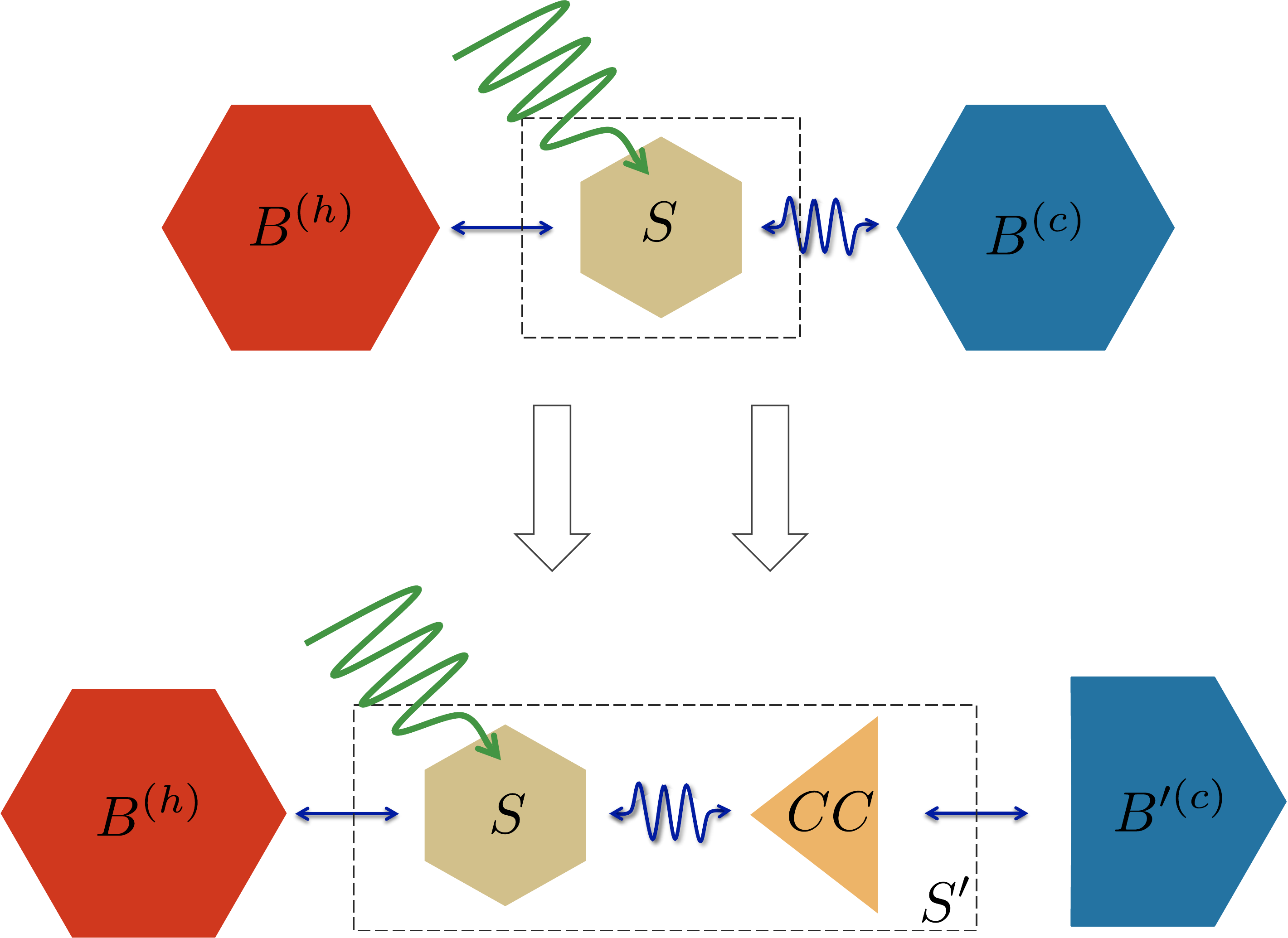}
\end{center}  
\caption{Sketch of the model before and after the CC mapping. A driven system $S$ coupled to a hot bath $B^{(h)}$ and a cold bath $B^{(c)}$ is 
mapped to a system coupled to the same hot bath and the CC, which is coupled to a residual bath $B'^{(c)}$. The oscillating arrows indicate the influence of the driving. The CC and the residual reservoir $B'^{(c)}$ are related by a unitary transformation to the original reservoir $B^{(c)}$.} 
\label{Fig:Model}
\end{figure}

The parameters $d_c$ and $d_h$ describe the overall coupling strength to each respective reservoir. The spectral density of the cold reservoir is peaked around resonance frequency $\omega_{res}$ and its structure can be tuned by the parameter $\gamma$, where the smaller $\gamma$ the more strongly peaked the spectral density is. The parameter $\omega_0$ is a reference frequency of the working medium. We will also consider only the coupling operator of the cold bath $S_{c}(t)$ to be time dependent. 
Figure~\ref{Fig:Model} (top panel) shows a sketch of the model.

\subsection{Collective Coordinate Mapping}\label{sec:CC}
To be able to capture non-Markovian and strong coupling effects of our model we follow the procedure discussed in 
\cite{Iles-Smith2014,Strasberg2016,Iles-Smith2016,Newman2017,Strasberg2017,Strasberg2017a,Schaller2017} and introduce a CC, also known as reaction coordinate, from the reservoir as part of the system. This is done by a unitary transformation applied to the bath degrees of freedom, which we will apply to the cold bath only because of its structured form. In the following we will refer to this method as the collective coordinate mapping.

The CC is defined as
\begin{equation}\label{eq:def_cc}
\sum\limits_k c_{kc} x_{kc} = \lambda_0 X_1  ,
\end{equation}
with $\lambda_0$ an unspecified parameter so far. After the mapping (see figure~\ref{Fig:Model}), we obtain
\begin{equation}\label{eq:H'}
H = H_{S'}(t) + H_{B'}^{(c)} + H_{S'B'}^{(c)} + H_{B}^{(h)} + H_{SB}^{(h)} \, ,
\end{equation}
with a redefined ``supersystem'' Hamiltonian 
\begin{equation}\label{eq:Hs'}
H_{S'}(t) = H_S(t) + \frac{1}{2}\left[P_1^2 + \omega_{CC}^2\left(X_1-\frac{\lambda_0}{\omega_{CC}^2}S_c(t)\right)^2\right],
\end{equation}
and residual cold reservoir $B'$ described by
\begin{equation}\label{eq:Hb'}
H_{B'}^{(c)} + H_{S'B'}^{(c)} = \frac{1}{2}\sum\limits_{k > 1} \left[P_k^2 + \Omega_k^2\left(X_k - \frac{C_k}{\Omega_k^2}X_1\right)^2\right].
\end{equation}
Here, $X_1$ and $P_1$ are the position and momentum operators of the CC and $X_k$ and $P_k$ position and momentum 
operators of the residual cold bath $B'$. The CC has natural frequency $\omega_{CC}$ and the spectral density of $B'$ 
is defined as $J_c'(\omega) \equiv \frac{\pi}{2}\sum\limits_{k> 1} \frac{C_k^2}{\Omega_k}\delta(\omega-\Omega_k)$. In the 
continuum limit, all new parameters and $J_c'(\omega)$ can be expressed in terms of the original 
spectral density $J_c(\omega)$ (see~\ref{App:CC} or \cite{Martinazzo2011}). For our choice of spectral density, equation~(\ref{eq:CSD}), we have 
\begin{equation}\label{eq:New_SD}
J'_c(\omega) = \gamma \omega, \quad \omega_{CC} = \omega_{res}, \quad \lambda_0 = d_c.
\end{equation}
Figure \ref{Fig:Model} shows a sketch of the original and mapped model with the newly defined supersystem $S'$.
In terms of creation and annihilation operators the new working medium supersystem Hamiltonian can be written as
\begin{equation}\label{eq:Super_H}
H_{S'}(t) = H_S(t) + \omega_{CC}\, a^{\dagger}a - \frac{\lambda_0}{\sqrt{2 \omega_{CC}}}\, S_{c}(t) \,(a + a^{\dagger}).
\end{equation}
One of the key features of the mapping is that an increase in the interaction strength between working medium and 
cold bath only increases the interaction between working medium and CC. The coupling strength between CC and residual 
bath is unaffected (see equation~\ref{eq:New_SD} or  \ref{App:CC}), allowing to treat the supersystem with standard 
weak coupling master equations when $\gamma$ can be considered small.

The CC mapping has been used to study an Otto cycle stroke type engine \cite{Newman2017}, stochastic thermodynamics based on coarse-graining \cite{Strasberg2017}, continuously coupled but not periodically driven engines \cite{Strasberg2016}, a fermionic autonomous Maxwell demon \cite{Strasberg2017a} and a fermionic electronic Maxwell demon \cite{Schaller2017}, all in the strong coupling and non-Markovian regime.  It has offered an accurate method for the study of open 
quantum systems apart from the thermodynamic applications \cite{Martinazzo2011a,Iles-Smith2014,Iles-Smith2016}
and is closely related to the method of  “time evolving density matrix using orthogonal polynomials algorithm” (TEDOPA) \cite{Prior2010,Chin2010,Woods2014,Rosenbach2016}. It was, however, not applied to the study of periodically driven systems so far. We point out that even though the CC mapping was applied here to the specific case of a spectral density with a single peak (see equation (\ref{eq:CSD})), spectral densities with many peaks or general non-Ohmic forms may be addressed with the mapping if the use of multiple collective coordinates is considered \cite{huh2014a}.

\subsection{Floquet Master Equation with Counting Field}\label{sec:Floquet}
We want to obtain a master equation for the time-periodic supersystem $S'$ that permits us to study its thermodynamic properties based on a rigorous framework. For this purpose we introduce a counting field $\chi_{\nu}$ for each reservoir as in \cite{Esposito2009,Cuetara2015} (see~\ref{App:Counting Field}) and apply Floquet theory for open systems (see~\ref{App:Floquet theory} and \ref{App: Floquet ME}).

Formally, the counting field $\chi_{\nu}$ is introduced by defining the modified density matrix
\begin{equation}
\rho_{tot}(\chi_{\nu}, t) \equiv   U(\chi_{\nu},t)\rho_{tot}(0) U^{\dagger}(-\chi_{\nu},t) , \label{eq:CF_def1}
\end{equation}
with total (supersystem plus reservoirs) initial density matrix $\rho_{tot}(0)$ and modified evolution operator $U(\chi_{\nu},t) = e^{-i \, \chi_{\nu} H_B^{(\nu)}/2} U(t) \, e^{ \,i \,  \chi_{\nu} H_B^{(\nu)}/2} $, where $U(t)$ is the evolution operator corresponding to the Hamiltonian of equation~(\ref{eq:H'}). The introduction of a counting field will allow us to obtain the change in energy of each reservoir from the operator $\rho(\chi_{\nu}, t)\equiv \text{Tr}_B\left\lbrace \rho_{tot}(\chi_{\nu}, t) \right\rbrace$ (see~\ref{App:Counting Field} for details).
To compute the reduced density operator dressed with counting field $\rho(\chi_{\nu}, t)$, a generalized master equation can be derived by performing the Born and Markov approximations to equation~(\ref{eq:CF_def1}). In the
Schr\"{o}dinger picture it has the form $\partial_t\rho(\chi_{\nu}, t) = \mathcal{L}(\chi_{\nu},t) \rho(\chi_{\nu},t)$, where 
$\mathcal{L}(\chi_{\nu},t)$ is a time periodic superoperator that has the same frequency as the driving. The master equation for the reduced density operator of the supersystem is then obtained by taking $\chi_{\nu} = 0$.
In general, $\rho(t)$ is not a time periodic function, but we are only interested in the dynamics in the long-time limit ($t\rightarrow \infty$). Here, we expect $\rho(t)$ to acquire the same periodicity as the driving, such that the master equation in the long-time limit may be written as
\begin{equation}\label{eq:Extended}
i \, n \,  \omega_L \rho_n = \sum_k \mathcal{L}_k \, \rho_{n-k} \, ,
\end{equation}
where $\rho_q$ and $\mathcal{L}_q$ are Fourier components defined by $\rho(t) = \sum_q e^{i   \omega_L q t} \rho_q$ and $\mathcal{L}(t) = \sum_q e^{i \omega_L q  t} \mathcal{L}_q$, respectively. The change in energy of the reservoir can then be obtained from (see~\ref{App:Counting Field})
\begin{equation} \label{eq:Heat_CF}
\left\langle \dot{H}_B^{(\nu)} \right\rangle  = \left.  i \frac{d}{dt}\frac{\partial}{\partial \chi_{\nu}} \text{Tr}\left\lbrace \rho(\chi_{\nu}, t) \right\rbrace  \right\vert_{\chi_{\nu}=0} = i\,  \text{Tr}\left\lbrace  \mathcal{L}' (0, t) \rho(t) \right\rbrace, 
\end{equation} 
where the prime indicates a derivative with respect to $\chi_{\nu}$. Although this formulation is associated with the statistics of two consecutive measurements \cite{Esposito2009}, the effect of the first measurement only shows up in high order moments and can be accounted for by means of a generalized fluctuation theorem \cite{Cerrillo2016}.

In the standard weak coupling approach, it is usually assumed that the system's level broadening is much smaller than its level spacing such that non-resonant terms may be neglected.  This is 
known as the secular approximation \cite{Breuer2002}, whose advantage is that the master equation generator is of Lindblad form. It also has the advantage of simplifying the study numerically since at steady state 
one only has to deal with a vector of populations and not the full density matrix. However, the secular approximation is problematic here because, due to the periodic time dependence of $H_{S'}(t)$, we resort to Floquet theory for 
the solution, where the dynamics of the supersystem is determined by its quasienergies (chosen to lie in the same 
Brillouin zone) and respective Floquet modes (see~\ref{App:Floquet theory}). Due to the high (infinite) dimension 
of the CC, there is a large amount of near-degenerancies in the chosen Brillouin zone, thus making the secular approximation in general not 
suitable for our analysis \cite{Hone2009}. 

\subsection{Steady state Thermodynamics} \label{sec:Thermo}
Due to the periodicity of the working medium, expectation values in the long-time limit will asymptotically oscillate as a function of time with period $T$. It is natural then to consider their average over one period, i.e.  $\overline{A} = \int_0^T dt \, A / T $.

The first law (energy balance) and second law (positivity of entropy production rate) in the long-time limit stipulate 
\begin{align}
\overline{\dot{Q}_c}  + \overline{\dot{Q}_h} + \overline{\dot{W}}  &= 0 \, ,  \label{eq:First_law} \\
- \beta_h \overline{\dot{Q}_h} -\beta_c \overline{\dot{Q}_c} &\geq 0 \,. \label{eq:Entropy_prod}
\end{align}
Due to the weak coupling between supersystem and reservoirs, we identify the change in energy of the reservoir, calculated using the counting field $\chi_{\nu}$, with the heat flow $\dot{Q}_{\nu}$ \cite{Strasberg2016}, defined to be positive if it flows into the system. For the hot bath and cold bath we define respectively
\begin{equation}
\dot{Q}_{h}(t) \equiv -\langle  \dot{H}_B^{(h)} \rangle, \qquad  \dot{Q}_{c}(t) \equiv -\langle  \dot{H}_{B'}^{(c)} \rangle.
\end{equation}
The rate of work $\dot{W}$ performed on or exerted by the system is then fixed by the first law (\ref{eq:First_law}). 
Negative values indicate that work is extracted. Positivity of the entropy production rate, equation~(\ref{eq:Entropy_prod}), can be shown from the definition of entropy production \cite{Esposito2010} and the existence of a periodic steady state in the long-time limit. Under the assumption of weak coupling between supersystem and reservoirs, our use of master equation (\ref{eq:Extended}) showed no violation of the second law at steady state for all presented calculations.

The relation between the change in energy of the residual cold reservoir $B'$ and the change in energy of the original cold
reservoir $B$ is given by the change in energy of the CC. To see this, note that in the CC mapping the reservoir 
Hamiltonian is mapped to three different terms (see equations~(\ref{eq:H'}), (\ref{eq:Hs'}) and~\ref{App:CC})
\begin{equation}
H_{B}^{(c)}  =   H_{B'}^{(c)}  +  H_{CC} + H_{CCB'}^{(c)},
\end{equation}
the CC Hamiltonian $H_{CC}$, the residual bath Hamiltonian $H_{B'}^{(c)}$  and the interaction between CC and residual reservoir $H_{CCB'}^{(c)}$.  As the coupling between CC and residual reservoir is weak, we can approximate the expectation value of the residual reservoir by  $ \langle  \dot{H}_{B'}^{(c)} \rangle \approx \langle  \dot{H}_{B}^{(c)} \rangle -  \langle  \dot{H}_{CC} \rangle $. In the long-time limit, if we average over one period, we have  $ \overline{\langle  \dot{H}_{CC} \rangle} =0$ since $\rho(t)$ is periodic. Hence $ \overline{\langle  \dot{H}_{B'}^{(c)} \rangle} \approx \overline{\langle  \dot{H}_{B}^{(c)} \rangle }$.

Note that the CC mapping allows for a thermodynamically consistent study of a time-dependent interaction $ S_{\nu}(t)$ between the system and the cold reservoir.  This is due to the fact that, after the mapping, the time-dependent coupling operator only appears in the supersystem Hamiltonian. The interaction with the residual reservoir has no time dependence (see Sec.~\ref{sec:CC} or \ref{App:CC}).

\section{Results}\label{sec:Results}
All previously presented ideas apply to the case of an arbitrary working medium 
described with Hamiltonian $H_S(t)$ and arbitrary coupling operators $S_{\nu}(t)$. We now focus specifically on the case where the working medium is a driven two level system
\begin{equation}
H_S(t) =\frac{ \omega_0 }{2}\sigma_z + g \, \cos(\omega_L t) \sigma_x \, .
\end{equation}
Here, $\sigma_i$ denotes a Pauli matrix, $\omega_0> 0$ is the energy splitting of the two-level system in the absence of driving, $g$ the driving strength and $\omega_L$ is the frequency of the driving responsible for work extraction and injection. The system coupling operators are
\begin{equation}
S_h = \sigma_x /\sqrt{2 \, \omega_0}, \qquad S_c(t) = \sigma_x \sin(\omega_L t) /\sqrt{ 2 \,  \omega_0}, \quad
\end{equation}
where we have included a time dependence on the cold reservoir coupling operator. 

The mapped working medium consist then of a driven two level system coupled with the CC via a time dependent interaction. This bears resemblance to the setup of laser cooling of trapped ions \cite{Cirac1992}, where an ion interacts with an electric field. The ion is approximated by a two-level system and its motional degree of freedom is quantized in a harmonic potential.  Assuming the ion is well confined, one ends up with an interaction Hamiltonian between the ion and the harmonic potential that is time-dependent in a similar fashion as the coupling operator $S_c(t)$ (see~\ref{App:Laser Cooling}). Cooling in this context refers to the preparation of the state of the harmonic mode in a low occupation number.  It is optimized around the resonance condition where the frequency of the laser and of the harmonic potential add up to the energy splitting of the two-level system. Based on the similarities of this setup and our mapped model we explore this resonance condition which for our model reads
\begin{equation}\label{eq:Resonance}
\omega_0 = \omega_{L} + \omega_{CC}.
\end{equation}
Importantly, the setup of laser cooling follows from an exact limit of our model by considering a weak amplitude driving $g \ll \omega_0$ and taking the limit $\gamma \rightarrow 0$. In this case, there is no coupling to the residual reservoir, we have a two level system (ion) coupled to the CC (vibrational mode of the ion) and one bath (see inset in figure~\ref{Fig:LC}). This limit will be considered  later in Sec.~\ref{sec:single_reservoir}. For now we keep $\gamma$ finite so that the supersystem is coupled to both reservoirs.

\subsection{Weak coupling}\label{sec:weak}
First, we will consider weak coupling between working medium and cold bath. In the mapped model this translates to a 
weak interaction $\lambda_0 = d_c$ between the two-level system and CC. It is important to note that even when the coupling is weak, the 
structured shape of the spectral density can have a strong effect on the behavior of the system invalidating a conventional master equation approach for the two-level system (also see~\ref{App:Benchmark} for benchmarks related to that question).
\begin{figure}[t!]
\begin{center}
  \includegraphics[width=0.49\linewidth]{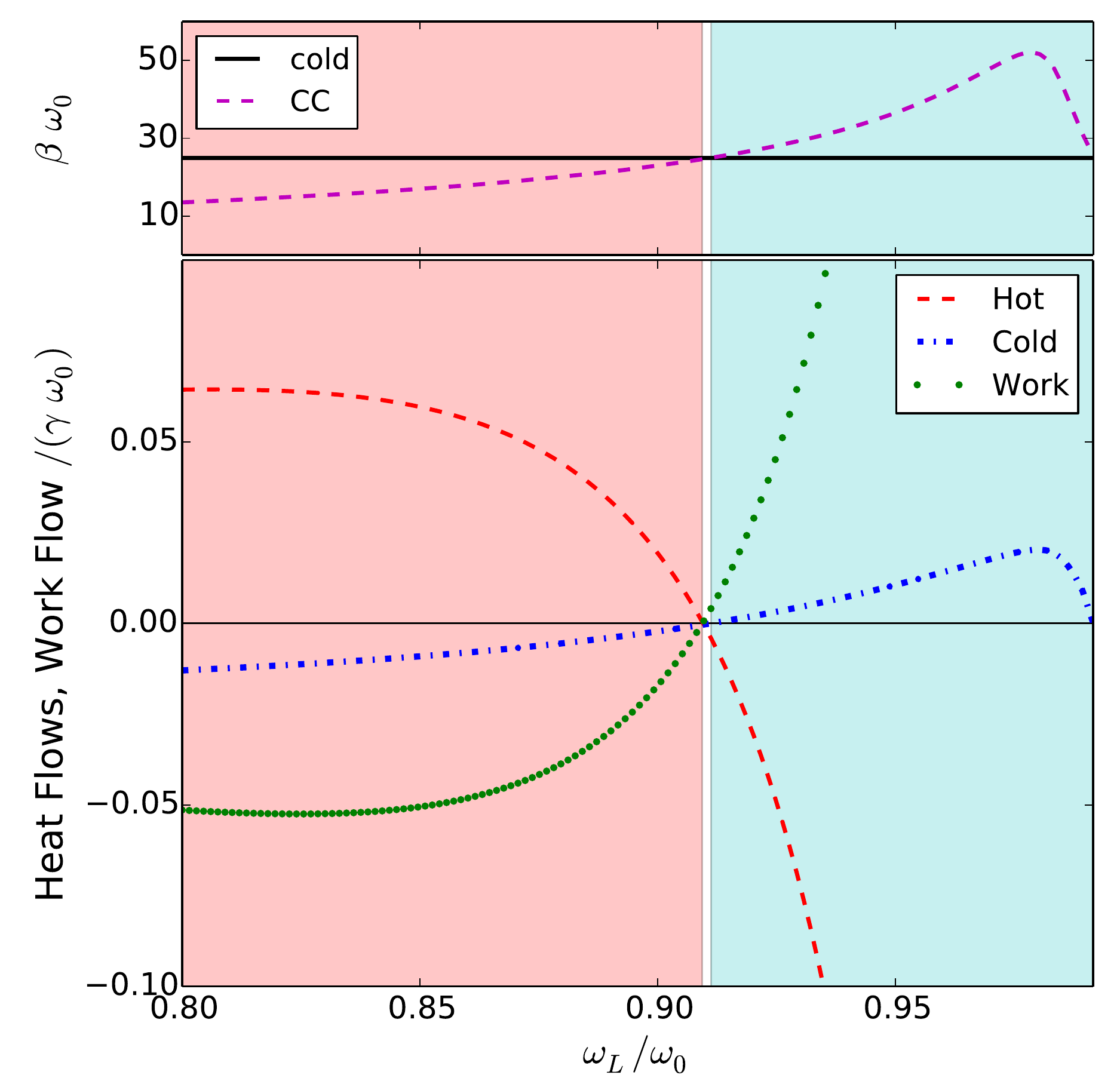}
\end{center}    
\caption{Bottom panel shows time averaged heat flows and work flows as a function of the driving frequency $\omega_L$ while keeping the resonance 
condition (\ref{eq:Resonance}). The red (left) shaded area corresponds to the region where the model 
behaves as a heat engine extracting work $(\dot{W} < 0)$. The blue (right) shaded area shows the region where the model behaves as a 
refrigerator cooling the cold bath $(\dot{Q}_c > 0)$.  Parameters used are: 
$g = 0.001 \omega_0,\, \gamma = 0.0004 \omega_0, \, d_c = 0.001 \omega_0^2, \, d_h = 0.005\omega_0^2, \, \beta_c\omega_0= 25,$ and $\beta_h \omega_0 = 2.22$. 
The top panel shows the effective inverse temperature of the CC (dashed line) as a function of the driving frequency, see equation~(\ref{eq:betaCC}) and the constant inverse temperature of the cold reservoir (continuous line) for the same parameters.}
\label{Fig:HE_Refri}
\end{figure}

Figure \ref{Fig:HE_Refri} (bottom panel) shows the heat flow to the two reservoirs and the work flow as a function of the driving 
frequency while the resonance condition is maintained. Two main regions can be identified in figure~\ref{Fig:HE_Refri} 
by looking at the conditions $\overline{\dot{W}}<0$ and $\overline{\dot{Q}_c} > 0$. When one of them is fulfilled the model either behaves 
as a heat engine or as a refrigerator. The first region is illustrated by the red (left) shaded area. Here, heat flows 
from the hot reservoir into the system $\overline{\dot{Q}_h}>0$, work is extracted $\overline{\dot{W}}<0$ and heat  flows from the system into 
the cold reservoir $\overline{\dot{Q}_c}<0$ making our model act as a heat engine. The blue shaded area shows the region where the 
model acts as a refrigerator. Here, work is applied to the system $\overline{\dot{W}}>0$, heat flows from the system to the hot 
reservoir $\overline{\dot{Q}_h}<0$ and from the cold reservoir to the system $\overline{\dot{Q}_c}>0$. We also note the existence of a small finite white region separating the transition between refrigerator and heat engine not seen in a previous study 
\cite{Gelbwaser-Klimovsky2013}, based on the standard Floquet-Markov approach, for a similar but simpler model.

\begin{figure}[t!]
\begin{center}
  \includegraphics[width=0.49\linewidth]{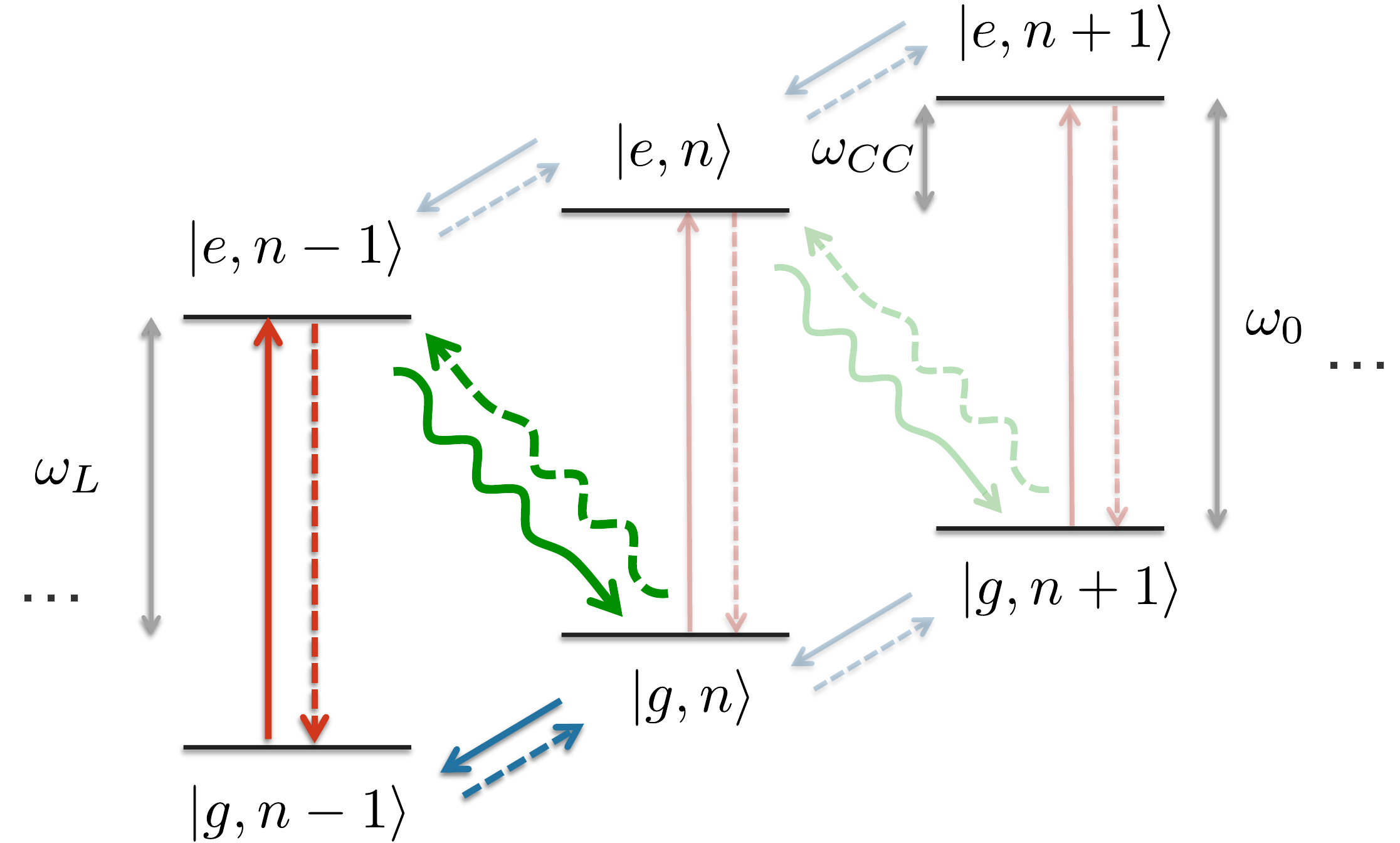}
 \end{center}
\caption{Mapped working medium acting as a refrigerator (dashed arrows) or as a heat engine (continuous arrows) at the resonance condition (\ref{eq:Resonance}). $\ket{e}$ and $\ket{g}$ refer to the excited and ground state of the two-level system. $\ket{n}$ refers to the Fock state $n$ of the CC.}
\label{Fig:Ener_level}
\end{figure}

To physically explain the heat engine and refrigerator regime, one can use figure~\ref{Fig:Ener_level}. It shows the energy levels of the mapped working medium [see equation~(\ref{eq:Super_H})] for very weak coupling and driving amplitude. In this parameter regime the cold reservoir mainly induces transitions between states of the CC, leaving the state of the two-level system unchanged. This is illustrated for example by the transitions $\ket{g,n} \leftrightarrow \ket{g, n-1}$ (blue arrows). On the other hand, the hot bath preferentially induces transitions that leave the Fock state of the CC fixed and flip the state of the two-level system as in $\ket{e,n-1}\leftrightarrow\ket{g, n-1}$ (red arrows). Under the chosen resonance condition, the driving part addresses transitions that alter both the state of the two-level system and the CC, and they are responsible for work extraction or injection. Such transitions are illustrated by e.g. $\ket{e,n-1}\leftrightarrow\ket{g, n}$ (green arrows). Fig \ref{Fig:Ener_level} only shows the most dominant transitions, additional transitions are induced by the reservoirs or the driving but they are suppressed due to the weak coupling or the resonance conditions and become less likely.

To further understand why our model behaves either as a refrigerator or a heat engine, let us consider the effective inverse temperature of the CC $\beta_{CC}$, implicitly defined by the equation 
\begin{equation}\label{eq:betaCC}
\overline{\langle a^{\dagger}a \rangle} =  \left[ \text{exp} \big( \beta_{CC}\,\omega_{CC} \big)   -1 \right]^{-1} \, , 
\end{equation}
where $\overline{\langle a^{\dagger}a \rangle}$ is the occupation number of the CC in the long-time limit averaged over a 
period. Whenever $\beta_{CC}$ is bigger than $\beta_{c}$ heat flows from the cold bath into supersystem $S'$ and increases 
the Fock state of the CC, therefore cooling the cold reservoir. This cooling is achieved by the injection of work 
simultaneously exciting the two-level system and lowering the state of the CC followed by heat flowing from the 
supersystem into the hot reservoir via a decay of the two-level system. This process is illustrated by the dashed 
arrows in figure~\ref{Fig:Ener_level}.  In the case that our model acts as a heat engine, heat flows from the hot 
reservoir to the two-level system and work is extracted by simultaneously exciting the CC and the two-level system 
decaying. Also, since now $\beta_{CC} < \beta_c$, heat no longer flows from the cold reservoir into the system. 
This process is indicated by the continuous arrows in figure~\ref{Fig:Ener_level}. The top panel of figure~\ref{Fig:HE_Refri} 
shows a plot of $\beta_{CC}$ as a function of the frequency. We see that the effective temperature of the CC (dashed line) and the temperature of the cold bath (continuous line) agree precisely at the boundary of the refrigerator region. We stress the fact that the discussion so far only applies for weak driving strength and weak coupling between working medium and cold bath, since otherwise the energy level diagram of figure~\ref{Fig:Ener_level} does not apply. The implicit definition of the effective temperature in equation (\ref{eq:betaCC}) is just a particular parametrisation. We have analysed the fluctuations around the average occupation number of the collective coordinate and found good agreement with those typical for a thermal state (see~\ref{App:Fluctuations}). In addition, we have checked explicitly that in this limit, the populations of the CC are close to a thermal distribution.

\subsection{Finite coupling strength} \label{sec:strong}
\begin{figure}[t!]
\begin{center}
  \includegraphics[width=0.49\linewidth]{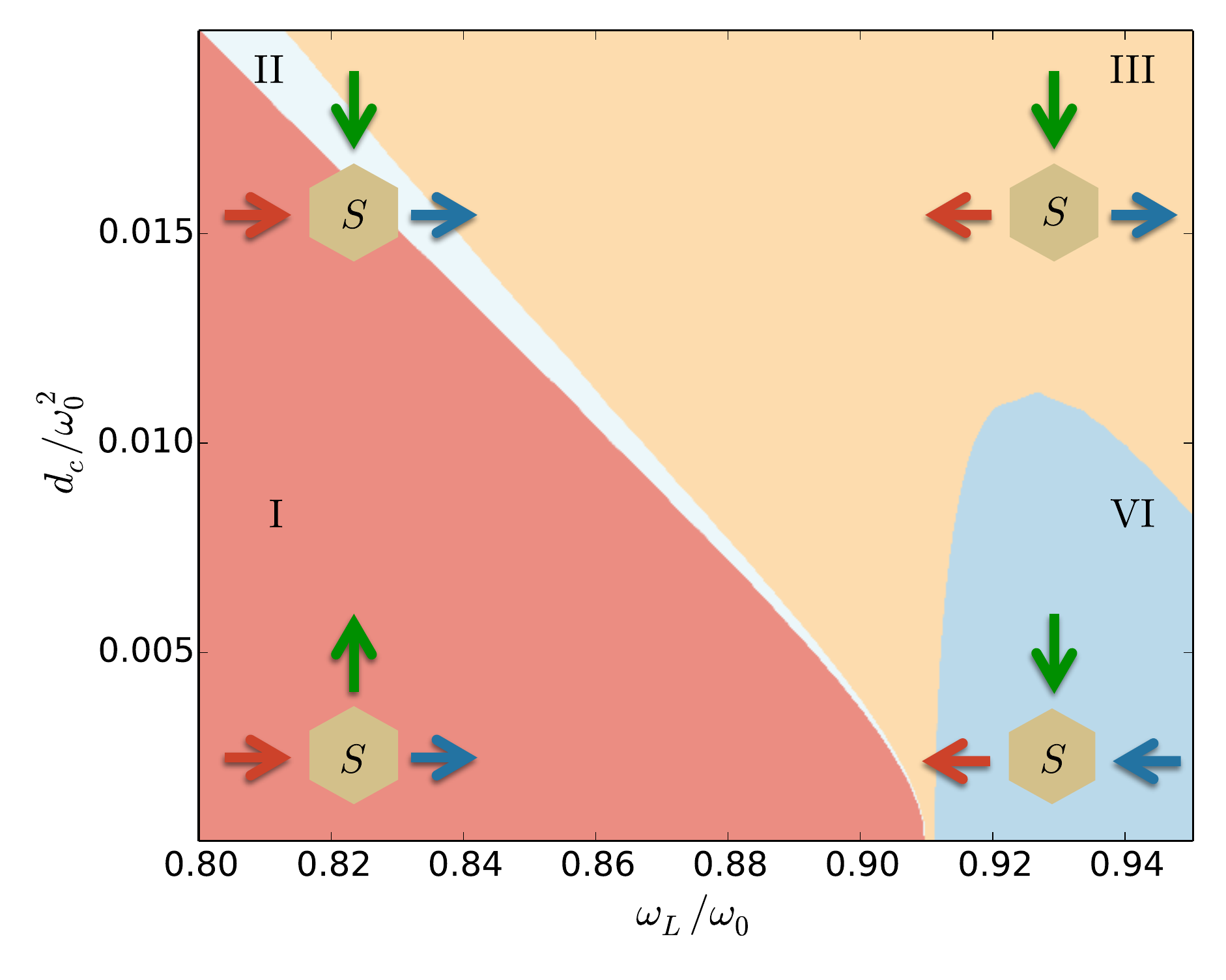}
  \includegraphics[width=0.49\linewidth]{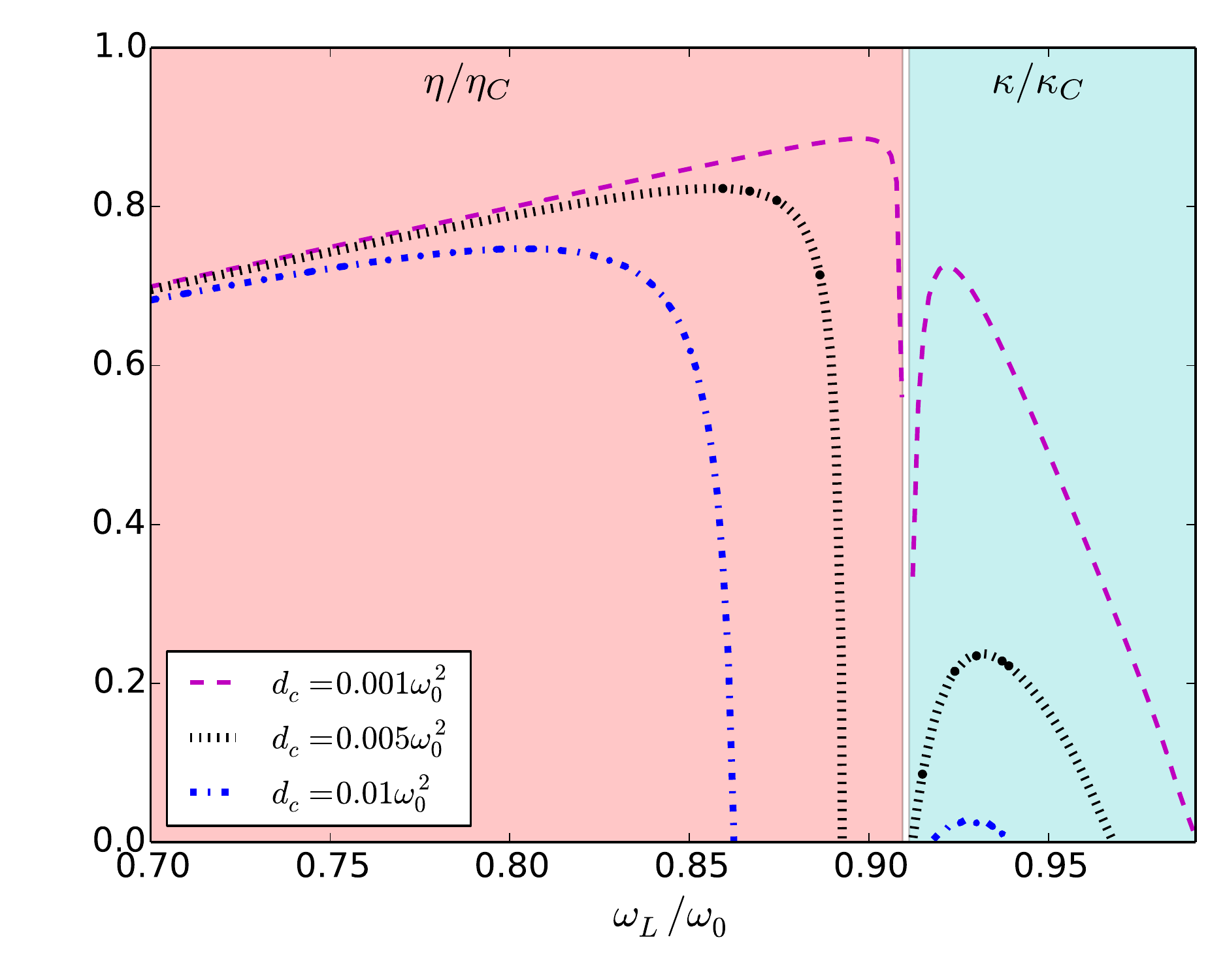}
\end{center}
\caption{Left: Phase diagram showing the different thermodynamic behaviors of the model. Parameters are the same as in figure~\ref{Fig:HE_Refri} except for the coupling $d_c$.  Vertical (green) arrows indicate whether work is performed on/by the system and horizontal arrows indicate the direction of the heat flow. Right: Efficiency and coefficient of performance normalized by the ideal Carnot bounds. Parameters are the same as in figure~\ref{Fig:HE_Refri} except for the coupling $d_c$. The shaded areas only apply for $d_c=0.001 \, \omega_0^2$.}
\label{Fig:Phase_D}
\end{figure}

As mentioned earlier, the transition between refrigerator and heat engine is not immediate, there exists a gap between 
the two regimes. Although already present in figure~\ref{Fig:HE_Refri}, we further explore this gap in the phase diagram of 
figure~\ref{Fig:Phase_D} (left) where four different regions can be identified depending on the coupling strength 
between working medium and cold reservoir. Regions I and IV indicate the regions previously introduced, where the model 
behaves as a heat engine or as a refrigerator, respectively. In both of the two middle regions (II and III) work is being done 
on the system and heat flows from the system into the cold reservoir. Regions II and III are distinguished by the direction 
of the heat flowing between the hot reservoir and the supersystem. For region III, heat flows from the system into the 
hot reservoir and for region II the opposite way. In region II work is applied to the system to assist the flow of heat 
from the hot reservoir to the cold reservoir and in region III work is converted into heat flowing into both reservoirs. 
The existence of these regions should not be ignored since their area grows as the coupling is increased. The refrigerator 
regime actually completely disappears. The immediate transition from a refrigerator to a heat engine as 
the one obtained in \cite{Gelbwaser-Klimovsky2013} only occurs in the ideal case of vanishing coupling strength.

\subsection{Performance} \label{sec:performance}
To quantify the performance of our model we introduce the efficiency of the heat engine $\eta$ and the coefficient of performance (COP) of the refrigerator $\kappa$. Both have ideal Carnot bounds given in terms of the bath temperatures. They are defined as:
\begin{align}
\eta = \frac{ -\dot{W} }{\dot{Q}_h}    \;  &\leq  \; \eta_{C} = \frac{\beta_c - \beta_h}{\beta_c}, \\
\kappa = \frac{\dot{Q}_c }{\dot{W}}     \;  &\leq  \;  \kappa_{C} = \frac{\beta_h}{\beta_c -\beta_h}.
\end{align}
Figure~\ref{Fig:Phase_D} (right) shows the efficiency and COP of our model for different coupling strengths. The 
shaded area only applies for the weakest coupling case $d_c = 0.001 \omega_0^{2} $. A clear decrease in performance can 
be seen as the coupling increases for both, heat engine and refrigerator. This is consistent with performance results 
obtained for the quantum Otto cycle \cite{Newman2017} and a continuously coupled but not driven three-level heat engine \cite{Strasberg2016}. We also see that, as the coupling decreases, the maximum in efficiency $\eta$ for the heat 
engine approaches the border between the white and red region, where the power goes to zero (see figure~\ref{Fig:HE_Refri}).

The exploration of stronger coupling regimes than the ones in figure~\ref{Fig:Phase_D} (left) requires a high level of truncation for the CC for convergence, making numerical simulations very demanding. This is also the case for a blue detuned resonance condition where $ \omega_L > \omega_0$, since the CC is heated increasing its occupation number, and no proper truncation of the Fock levels may be defined. In all presented results the dimension of the CC was considered finite and truncated at a level where convergent results were obtained.

\subsection{Single Reservoir: Laser cooling limit}\label{sec:single_reservoir}
\begin{figure}[t!]
\begin{center}
  \includegraphics[width=0.49\linewidth]{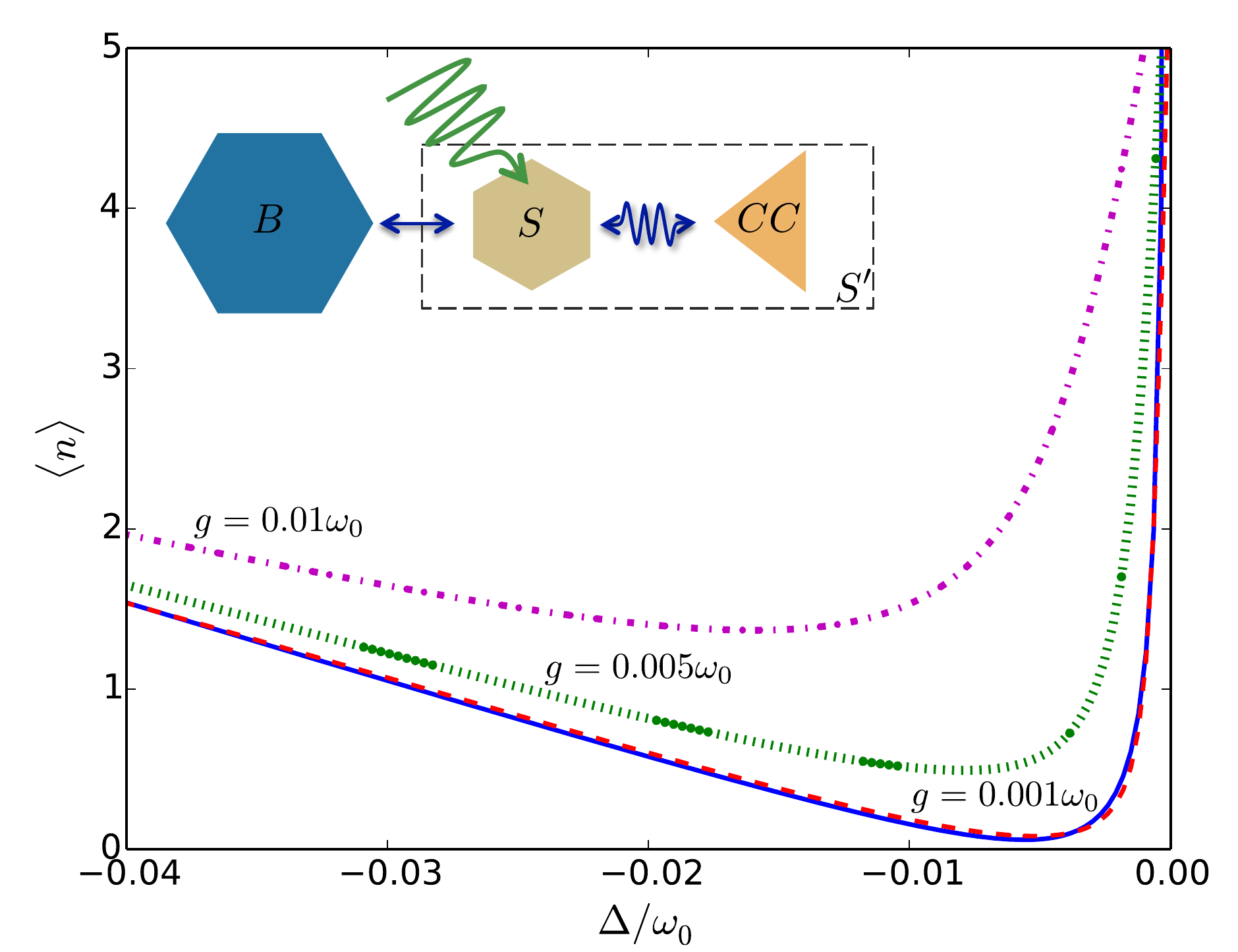}
\end{center}
\caption{Occupation number of CC as a function of the detuning between the two-level system and the driving frequency. 
Continuous line is obtained from the analytic results in \cite{Cirac1992}. Parameters used are: 
$d_h=0.005\omega_0^2 , \, d_c = 0.000113\omega_0^2, \, \beta \omega_0 = 10^4$ and $ \omega_{res}=0.005 \omega_0,$.}
\label{Fig:LC}
\end{figure}

We now consider the case of vanishing coupling between supersystem and residual bath $B'$: $\gamma \rightarrow 0$. This means that the bath is exclusively represented by the CC, and therefore there is no need to treat it perturbatively. The supersystem is then coupled only to a single reservoir. The setup for laser cooling of trapped ions \cite{Cirac1992} is well represented by this model (see~\ref{App:Laser Cooling}) if one thinks of the CC as the vibrational mode of an ion. The goal of such experiments is the preparation of the harmonic mode in its ground state, and the success is usually measured in terms of a low occupation number of the vibrational mode $\langle n \rangle$. The inset of figure~\ref{Fig:LC} shows a sketch of this set up.
Figure \ref{Fig:LC} shows the occupation number of the CC as a function of the detuning $\Delta = \omega_L - \omega_0$ for different driving amplitudes. We see that the optimal preparation (maximum cooling) is obtained for small driving amplitudes and around the resonance condition studied previously $\Delta = -\omega_{CC}$.  The continuous line is obtained from the analytical results in \cite{Cirac1992} where they showed that at steady state the occupation number is given by $\langle n \rangle = \frac{A_+}{A_- -A_+}$, with $ A_{\pm} = \Gamma f(\Delta \pm \nu)$ and $f(\Delta)= \frac{(\Gamma / 2)^2}{(\Gamma / 2)^2+\Delta^2}$, $\nu$ the frequency of the oscillator and $\Gamma$ the decay rate of the two-level system (ion). In figure~\ref{Fig:LC} we take $\nu = \omega_{CC}$ and $\Gamma = d_h$. It is important to state that the analytical results in \cite{Cirac1992} are obtained under approximations that we do not perform here such as the rotating wave approximation and the adiabatic elimination of the internal degrees of freedom of the ion. We have also only considered the case of low amplitude driving such that  scattering effects, as the recoil of the ion in the emission of a photon, can be neglected.

\section{Conclusions}\label{sec:Conclu}
We have provided a novel framework to study periodically driven thermal machines beyond the weak coupling and Markovian approximations. 
The CC mapping was used to overcome the loss of separability that appears for systems as their interaction strength with the bath increases, Floquet theory allowed us to accurately treat periodic time dependencies of working media and, with the use of full counting statistics, we were able to consistently perform a steady state thermodynamic analysis.

We specifically considered as the working medium a driven two-level system with time-dependent interaction to one of its reservoirs and dealt with the case where only this heat reservoir had a structured spectral density. We focus on the limit of a strongly peaked spectral density since that also guaranteed a weak coupling to the residual bath such that a master equation could be used. In the weak coupling and non-Markovian scenario, we confirmed the existence of two main regimes where the thermal machine works either as a refrigerator or a heat engine and gave an intuitive physical explanation of the observed difference and transition between these two regimes in terms of the occupation number of the CC and its effective temperature. However, we also additionally identified a small gap between both operational regimes.

Upon increasing the interaction strength between working medium and the cold reservoir, we observed an enlargement of the gap between the refrigerator and heat engine regime. We identified four different operation regimes for our model and witnessed the eventual disappearance of the refrigerator regime as the coupling increased.  In terms of efficiency, both the refrigerator and the heat engine showed a monotonic decrease of performance as a function of the coupling strength. Similar performance results for non periodically driven systems were observed in \cite{Newman2017, Strasberg2016} and higher efficiency in the weak coupling but non-Markovian regime was observed in \cite{Strasberg2016}. Nonetheless, as it was the case here, these results were focused on particular models, and the general question of whether strong coupling and non-Markovian effects are (un)favorable for the performance of quantum thermal machines is still open.
Strong coupling corrections to the second law due to the interaction strength were derived in \cite{Perarnau-Llobet2017}, also showing a lower performance, but restricted to thermodynamic protocols where the system is adiabatically driven and never simultaneously coupled to more than one thermal bath.

We also considered the case where the inclusion of the CC is enough to capture all the effects of one of the two 
baths, leaving the supersystem coupled only to one bath. By doing this we recovered the setup of laser cooling of trapped ions. The predicted occupation of the harmonic mode (CC) was shown to be in good agreement with the analytical results of \cite{Cirac1992}, confirming the adequacy of the proposed method.

Finally, we emphasize that the method presented here applies in general to any periodically driven thermal machine coupled linearly with a bosonic bath, so that more complex models may be addressed in the future.

\ack
S. R. acknowledges fruitful discussions with C. W. W\"achtler. The authors acknowledge financial support through DFG Grants No. GRK 1558, SFB 910 and BR 1928/9-1. P.S. acknowledges financial support by the European Research Council project NanoThermo (ERC-2015-CoG Agreement No. 681456).

\appendix

\section{Collective Coordinate Mapping}\label{App:CC}
 In the following it will be assumed that the system (supersystem) is in contact only with one bath, but additional reservoirs can be incorporated additively.
The relevant quantities of the mapped Hamiltonian are given in terms of the original spectral density 
\cite{Martinazzo2011} as
\begin{equation}
\delta \Omega_0^2 = \frac{2}{\pi}\int\limits_0^\infty d\omega \frac{J_0(\omega)}{\omega}, \qquad \lambda_0^2 = \frac{2}{\pi}\int\limits_0^\infty d\omega~\omega J_0(\omega),
\end{equation}
with $J_0(\omega) = J_c(\omega)$ and frequency of CC $\omega_{CC} = \frac{\lambda^2}{\delta \Omega_0^2}$. The residual bath $B'^{(c)}$ has a new spectral density $J_1(\omega) = J'_c(\omega)=\frac{\pi}{2}\sum\limits_k \frac{C_k^2}{\Omega_k}\delta(\omega-\Omega_k)$ which can be related to the original spectral density by
\begin{equation}
J_1(\omega) = \frac{\lambda_0^2 J_0(\omega)}{\left|W_0^+ (\omega)\right|^2} = \frac{\lambda_0^2 J_0(\omega)}{ \left( \frac{1}{\pi} \mathcal{P} \int_{-\infty}^{\infty} d\omega' \frac{J_0(\omega')}{\omega'-\omega}\right)^2   +  ( J_0(\omega))^2}, 
\end{equation}
where $W_0(z) = \frac{1}{\pi}\int\limits_{-\infty}^\infty d\omega~\frac{J_0(\omega)}{\omega -z}$, 
$W_0^+(\omega) = \lim \limits_{\varepsilon \searrow 0}W_0(\omega + i\varepsilon)$ and $\mathcal{P}$ indicates the principal value. Note that $J_0(\omega)$ is extended to negative values of $\omega$ via $J_0(-\omega) = -J_0(\omega)$. 
From these expressions it is easy to see that increasing the coupling between system and reservoir by 
$J_0(\omega) \rightarrow \alpha J_0(\omega)$ (with $\alpha\neq 1$) has no effect on the coupling between supersystem and 
residual reservoir since the multiplicative factor $\alpha$ cancels out in $J_1(\omega)$. 

It is important to note that the CC mapping can actually be applied more than once. If one wishes to apply the CC mapping recursively, it is then necessary that the behaviour of the spectral density $J_n(\omega)$ as $\omega \rightarrow \infty$, guarantees convergence of expressions for the  residual reservoir with spectral density $J_{n+1}(\omega)$. The index $n=0$ indicates the original spectral density of the bath. This can be guaranteed from the start by introducing a hard cutoff in the original spectral density $J_0(\omega)$ at a sufficient high frequency $\omega_R$. Here we apply the CC mapping only once.

For our particular model [see equation~(\ref{eq:CSD})], an analytical expression can be obtained for $J'_c(\omega)$. For $4 \omega_{res}^2 > \gamma^2$ it follows that \cite{Garg1985}
\begin{equation}
J'_c(\omega) = \gamma \omega.
\end{equation}
Additionally, we have $\lambda_0 = d_c$ and $\delta \Omega_0 = d_c / \omega_{res}$ such that $\omega_{CC} = \frac{\lambda^2}{\delta \Omega_0^2} = \omega_{res} $. In terms of creation and annihilation operators defined by $X_1 = \frac{1}{\omega_{CC}} (a+a^{\dagger})$, $P_1 = i \sqrt{\frac{\omega_{CC}}{2}} (a+a^{\dagger})$, $X_k = \frac{1}{2\Omega_k} (b_k+b_k^{\dagger})$ and $P_k = i \sqrt{\frac{\Omega_k}{2}} (b_k+b_k^{\dagger})$ equations~(\ref{eq:Hs'}) and (\ref{eq:Hb'}) can be written as
\begin{align}
H_{S'}(t) &= H_S(t) + \omega_{CC}\, a^{\dagger}a - \frac{\lambda_0}{\sqrt{2 \omega_{CC} }}\, S_{c}(t) \,(a + a^{\dagger}), \, \\
H_{B'}^{(c)} + H_{S'B'}^{(c)} &= \sum_k \Omega_k b_k^{\dagger} b_k + (a + a^{\dagger}) \,\sum_k \frac{C_k}{\sqrt{2 \Omega_k}} (b_k + b_k^{\dagger}),
\end{align}
where terms proportional to the identity and of second order in $C_k$ have been disregarded since we consider the case where the coupling to the residual reservoir is weak. Now the total Hamiltonian can be written as $H= H_S(t) + H_{CC} + H_{SCC} + H_{B'}^{(c)} + H_{CCB'}^{(c)} + H_{B}^{(h)} + H_{SB}^{(h)}$ with
\begin{align}
H_{CC} = \omega_{CC} a^{\dagger} a, &\qquad H_{SCC}= - \frac{\lambda_0}{\sqrt{2 \omega_{CC} }}\, S_{c}(t) \,(a + a^{\dagger}), \\
H_{B'}^{(c)} = \sum_k \Omega_k b_k^{\dagger} b_k , &\qquad H_{CCB'}^{(c)}  =  (a + a^{\dagger}) \,\sum_k \frac{C_k}{\sqrt{2 \Omega_k}} (b_k + b_k^{\dagger}).
\end{align}
Comparing with equation~(\ref{eq:H_xp}) we have $H_B^{(c)} = H_{CC} + H_{B'}^{(c)} + H_{CCB'}^{(c)}$.

\section{Counting Field} \label{App:Counting Field}
In order to study the thermodynamics of our system we follow the full counting statistics formalism reviewed
in \cite{Esposito2009}. The total Hamiltonian, including system and reservoir is $H=H_S(t) + H_B + H_{SB}$ and we assume an initial factorizing state of the form $\rho_{tot}(0) = \rho(0) \otimes \rho_B $, with $\rho_B \sim e^{-\beta H_B}$. Note that we have considered the interaction Hamiltonian $H_{SB}$ to be time-independent. This term has a periodic time dependence in our model but after the CC mapping this time dependence only appears in the supersystem Hamiltonian and not in the interaction to the residual bath. What is presented in this section also applies for the case of replacing $S$ by $S'$ and $B $ by $B'$.  Let us define the  modified density matrix
\begin{equation}
\rho_{tot}(\chi, t) \equiv   U(\chi,t)\rho_{tot}(0) U^{\dagger}(-\chi,t) , \label{eq:CF_def}
\end{equation}
with total (supersystem plus reservoirs) initial density matrix $\rho_{tot}(0)$ and modified evolution operator $U(\chi,t) = e^{-i \, \chi H_B/2} U(t) \, e^{ \,i \,  \chi H_B/2} $, where $U(t)$ is the evolution operator corresponding to Hamiltonian H.
 The variable $\chi$ is usually referred to as counting field. The evolution of operator $\rho_{tot}(\chi, t)$  is given by 
\begin{equation}
\partial_t \rho_{tot}(\chi, t) = -i \left[   H(\chi,t) \rho_{tot}(\chi,t) - \rho_{tot}(\chi,t)H(-\chi,t) \right] . \label{partial_total_rho}
\end{equation} 
Taking the trace over the reservoir degrees of freedom we define
\begin{equation}
 \rho(\chi, t) \equiv \text{Tr}_B\left\lbrace \rho_{tot}(\chi, t) \right\rbrace .
\end{equation} 
Note that the total density matrix and the reduced density matrix of our system are both recovered by setting $\chi=0$. 
The moment generating function \cite{Esposito2009} associated with the probability $p(\Delta E)$ of projectively measuring $H_B$ at time $t$ obtaining $E_t$ and at time $0$ obtaining $E_0$ is
\begin{equation}
G(\chi) = \int d\Delta E \, e^{-i \chi \Delta E } p(\Delta E) = \text{Tr}\left\lbrace \rho(\chi, t) \right\rbrace ,
\end{equation} 
with $\Delta E = E_t-E_0 $. It allows us to obtain the statistics of the energy transferred between system and reservoir by simple differentiation 
\begin{equation}
\left\langle \Delta E^n \right\rangle  =  \left. - \frac{\partial^n}{\partial (i \chi)^ n} G(\chi) \right\vert_{\chi=0}.
\end{equation} 
Note that for a steady-state thermodynamic analysis we are only interested in the change in time of the energy of the reservoir given by equation~(\ref{eq:Heat_CF}).

\section{Floquet Theory and Extended Space} \label{App:Floquet theory}
Floquet's theorem establishes that, for a time-periodic Hamiltonian $H_S(t) = \sum_k e^{ i k \,\omega_L t} H_k$, with period $T=\frac{2 \pi}{\omega_L}$, a solution to Schr\"odinger's equation is given by $ \ket{\psi_r(t)} = e^{- i \varepsilon_r t} \ket{r(t)}$, where $\varepsilon_r$ are called quasienergies and $\ket{r(t)}$ Floquet modes (states). The Floquet modes are time periodic and form a complete basis. To find them one solves the eigenvalue problem
\begin{equation}\label{Flo_eigenvalue}
\left(  H_S(t) - i \partial_t       \right)  \ket{r(t)} = \varepsilon_r \ket{r(t)}.
\end{equation}
The periodicity of the Floquet modes allows us to map the eigenvalue problem (\ref{Flo_eigenvalue}) to a time independent one in an extended Hilbert space, sometimes referred to as Sambe space \cite{Sambe1973}. This is done by introducing an infinite dimensional space with integer quantum numbers. Its basis is given by
\begin{equation}
B_{T_L} = \left\lbrace \ldots, \ket{-3} ,\ket{-2}  ,\ket{-1}  ,\ket{0}  ,\ket{1}  ,\ket{2} , \ket{3}            \ldots      \right\rbrace. 
\end{equation}
In this space we define the operators $F_k$ and $F_z$ by their action on a state from $B_{T_L}$. 
\begin{equation}
F_k\ket{m} = \ket{m+k}, \qquad F_z \ket{m} = m \ket{m}. 
\end{equation}
These operators will help us write equation~(\ref{Flo_eigenvalue}) in a simpler time independent form. We also have a basis for our quantum system living in Hilbert space $\mathcal{H}_S$. Its basis is denoted by $B_S = \left\lbrace \ket{\phi_1} ,\ket{\phi_2}  ,\ket{\phi_3}   \ldots \ket{\phi_K}   \right\rbrace$. Now, we can build the basis of the extended Hilbert space, which is the space in which our eigenvalue problem will become time independent. 
\begin{equation}
B =  B_{T_L}\otimes B_S = \big\{ \ket{n,\phi} \rangle \, |n\in\mathbb{Z},i\in\{1,\dots,K\}\big\} .
\end{equation}
Vectors in the extended Hilbert space will be denoted by a double ket notation $\ket{r}\rangle$. Its corresponding state at time $t$ in  $\mathcal{H}_S$ will be denoted by $\ket{r(t)}$. Conversely, a periodic state $\ket{u(t)}=\ket{u(t+T)}$ is denoted in the extended Hilbert space by $\ket{u}\rangle$. 
If an operator has a periodic time dependence $O(t)= \sum_k O_k e^{i \omega_L t}$, its form in extended space is defined as
\begin{equation}
O_{ext} = \sum_k F_k \otimes O_k.
\end{equation}
Furthermore, the scalar product in the extended space is defined by
\begin{equation}
\langle \braket{u}{v}\rangle = \frac{1}{T} \int_0^T dt \braket{u(t)}{v(t)}= \overline{\braket{u(t)}{v(t)}}.
\end{equation}
With the previous definitions it is now possible to write  the operator Q = $H(t)-i \partial_t$  and eigenvalue problem (\ref{Flo_eigenvalue}) in extended space as
\begin{equation}
Q_{ext} =\sum_k F_k \otimes H_k + \omega_L F_z \otimes \mathbb{1} , \qquad \longrightarrow \qquad Q_{ext} \ket{r_m}\rangle = \varepsilon _{rm} \ket{r_m}\rangle,
\end{equation}
where $H_k$ are the Fourier components of the Hamiltonian $H(t)$. The reason why the eigenstates and quasienergies have 
been denoted with an additional index $m$ in the extended space is because they actually carry redundant information. 
This becomes apparent  when one considers a solution of (\ref{Flo_eigenvalue}) and defines $\varepsilon_{rm}=\varepsilon_r + m \omega_L$ 
and $\ket{r_m(t)}=\ket{r(t)} e^{i m \omega_L t}$ such that $\ket{\psi_r(t)} = e^{- i \varepsilon_r t} \ket{r(t)}= e^{- i \varepsilon_rm t} \ket{r_m(t)}$. 
This means that the state $\ket{\psi_r(t)}$ can actually be constructed from both solutions $\ket{r_m}\rangle$ or $\ket{r_{m'}}\rangle$. To completely 
characterize the problem one has to choose the Floquet modes in the extended space whose quasienergies lie in the same Brillouin zone. 
From now on we will denote Floquet modes in the extended space just by $\ket{r}\rangle$, assuming that all lie in the same Brillouin zone.\\

With the Floquet modes at hand, one can find the evolution of any operator. Let us consider an arbitrary operator $S$.
\begin{equation}\label{S_decomp}
S(t) = U^{\dagger}_S(t) S U_S(t) = \sum_{\omega, n }  e^{i (\omega + n \omega_L)  t} \, S_{\omega, n}, 
\end{equation}
\noindent where $S_{\omega, n} = \left[ \int_0^T \frac{dt}{T}  \bra{r(t)} S \, e^{-i n \omega_L t} \ket{r'(t)}   \right]\, \ket{r(0)}  \bra{r'(0)}$ and $\varepsilon_r - \varepsilon_{r'}   = \omega$. Depending on the form of $\ket{r(t)}$ and $S$, calculating the integral in square brackets might not be trivial. In extended space this is easily calculated just by
\begin{equation}
\int_0^T \frac{dt}{T}  \bra{r(t)} S \, e^{-i n \omega_L t} \ket{r'(t)} = \langle \bra{r} F_{-n}\otimes S \ket{r'} \rangle.
\end{equation}
With decomposition (\ref{S_decomp}) it is now straight forward to obtain a master equation for a driven open quantum system. For a more complete study and review of Floquet theory and driven systems the reader is referred to \cite{Grifoni1998, Eckardt2015}.

\section{Floquet Master Equation} \label{App: Floquet ME}
Starting from equation~(\ref{partial_total_rho}), going to the interaction picture and performing the standard Born and Markov approximations \cite{Breuer2002} we obtain  
 \begin{equation}
 \begin{aligned}
 \partial_t \tilde{\rho}(\chi, t) = 
  &- \int^{\infty}_0 ds \text{Tr}_B \lbrace  \tilde{H}_I(\chi,t) \tilde{H}_I(\chi,t-s) \tilde{\rho}(\chi, t) \rho_B   
 - \tilde{H}_I(\chi,t) \tilde{\rho}(\chi, t) \rho_B \tilde{H}_I(-\chi,t-s)  \\
 &-   \tilde{H}_I(\chi,t-s) \tilde{\rho}(\chi, t) \rho_B \tilde{H}_I(-\chi,t)  
 +   \tilde{\rho}(\chi, t) \rho_B \tilde{H}_I(-\chi,t-s) \tilde{H}_I(-\chi,t)  \rbrace. 
 \end{aligned}
 \end{equation}
The interaction picture is defined by $\tilde{A}(t) = U_0^{\dagger}(t)A U_0(t) $, with $U_0(t)$ the evolution operator associated to Hamiltonian $H_0(t)=H_S(t)+H_B$. We take the interaction Hamiltonian to have the form $H_{SB} = S \otimes B=S \otimes \sum_k c_k x_k$ and define the correlation function $C(\chi, t) \equiv \langle \tilde{B}(\chi,t)B \rangle = \text{Tr}_B\left\lbrace   \tilde{B}(\chi,t)B  \rho_B \right\rbrace  $. Using the fact that $\left\langle \tilde{B}(\chi,t) \tilde{B}(\xi, s)    \right\rangle  = \left\langle \tilde{B}(\chi-\xi,t-s) B \right\rangle$, we have
 \begin{equation}
  \begin{aligned}
 \partial_t \tilde{\rho}(\chi, t) = 
  &- \int^{\infty}_0 ds  \lbrace  C(0,s)  \tilde{S}(t)\tilde{S}(t-s) \tilde{\rho}(\chi, t)   
 -      C(-2\chi,-s)  \tilde{S}(t)  \tilde{\rho}(\chi, t) \tilde{S}(t-s) \\
 &-   C(-2\chi,s)   \tilde{S}(t-s) \tilde{\rho}(\chi, t)  \tilde{S}(t)
 +  C(0,-s)   \tilde{\rho}(\chi, t) \tilde{S}(t-s) \tilde{S}(t)\rbrace. 
  \end{aligned}
 \end{equation} 
The correlation functions can actually be written in terms of the spectral density 
$J(\omega) \equiv \frac{\pi}{2}\sum\limits_k \frac{c_{k}^2}{\omega_{k}}\delta(\omega - \omega_{k})$ of the reservoir as $ C(-2\chi,t) = \frac{1}{\pi} \int^{\infty}_{\infty} d \omega e^{- i \omega (\chi + t)} J(\omega) \left[ 1 + N_\omega \right] $, 
with the Bose-Einstein distribution $N(\omega) = \frac{1}{e^{\beta \omega} -1 }$. 

The periodicity of $H_S(t)$ allows us to decompose the system operators in the interaction picture as in equation~(\ref{S_decomp}). Performing the integrals over $s$ and $\omega$ with the help of $\int^{\infty}_0  ds \, e^{i \omega s} = \pi \delta(\omega) + i \, \mathcal{P}  \, \frac{1}{\omega}$ and disregarding the principal value $\mathcal{P} $ term, one ends up with an equation in the Schr\"odinger picture of the form
\begin{equation} \label{ME_CF}
 \begin{aligned}
\partial_t \rho(\chi, t) &= \mathcal{L}(\chi,t) \rho(\chi,t)   \\
= &- \, i \left[   H_S(t), \rho(\chi,t) \right]   \\ 
& - \sum_{\left\lbrace \omega,n\right\rbrace } e^{i n \omega_L t }  \lbrace   J(\Delta_{\omega,n}) N(\Delta_{\omega,n}) S S_{\omega,n}(t)  \rho(\chi, t) \\
 & -  J(\Delta_{\omega,n})  \left[1+ N(\Delta_{\omega,n}) \right]   S  \rho(\chi, t) S_{\omega,n}(t) e^{i \Delta_{\omega,n} \chi}   \\ 
 & -  J(\Delta_{\omega,n})  N(\Delta_{\omega,n}) S_{\omega,n}(t) \rho(\chi, t)  S e^{-i \Delta_{\omega,n} \chi} \\ 
 &  +  J(\Delta_{\omega,n})  \left[1+ N(\Delta_{\omega,n}) \right] \rho(\chi, t) S_{\omega,n}(t) S\rbrace, 
 \end{aligned}
\end{equation}
with $\Delta_{\omega,n} = \omega + n \omega_L$ and 
$S_{\omega,n}(t) = \left[ \int_0^T \frac{dt}{T}  \bra{r(t)} S \, e^{-i n \omega_L t} \ket{r'(t)}   \right]\, \ket{r(t)}  \bra{r'(t)} $ 
such that $\omega = \varepsilon_r - \varepsilon_{r'}$. 
Note that due to the periodicity of the Floquet modes $\ket{r(t)}$, the superoperator $\mathcal{L}(\chi,t)$ also has the same periodicity. The heat flow is obtained by (see Sec.~\ref{sec:Thermo} and \ref{App:Counting Field}) 
\begin{equation} 
 \begin{aligned}
\dot{Q}(t) &= -\partial_t \left\langle  H_B \right\rangle  = \text{Tr}\left\lbrace \left(  \partial_{i\chi}  \mathcal{L}(\chi, t) \right)  \cdot \rho(\chi, t)     \right\rbrace \vert_{\chi=0}  \\
&= \sum_{\left\lbrace \omega,n\right\rbrace } e^{i n\omega_L t }  \quad  \text{Tr}    \lbrace  -  J(\Delta_{\omega,n})  \left[1+ N(\Delta_{\omega,n}) \right]\Delta_{\omega,n} S \rho(t) S_{\omega,n}(t) \\ 
& \qquad + J(\Delta_{\omega,n})  N(\Delta_{\omega,n}) \Delta_{\omega,n} S_{\omega,n}(t) \rho(t) S  \rbrace.
 \end{aligned} 
\end{equation}
Finally the evolution of the system can be computed just by setting $\chi = 0$ in equation~(\ref{ME_CF}),
\begin{equation} \label{ME_schr}
 \begin{aligned}
\partial_t \rho(t) = &- \, i \left[ H(t), \rho(t)    \right] - \sum_{\left\lbrace \omega,n\right\rbrace } e^{i n\omega_L   t }   J(\Delta_{\omega,n})  \lbrace     N(\Delta_{\omega,n}) 
\left[    S S_{\omega,n}(t) \rho(t)     -  S_{\omega,n}(t) \rho(t) S   \right] \\
& + \left[1+ N(\Delta_{\omega,n}) \right]   \left[    \rho(t)S_{\omega,n}(t) S   - S \rho(t)S_{\omega,n}(t)     \right]  \rbrace . 
 \end{aligned}
\end{equation} 
Assuming that in the long-time limit the density matrix $\rho(t)$ is time periodic with the same period as the Floquet modes we obtain equation~(\ref{eq:Extended}).  In the extended space this equation has the form
\begin{equation}
 \left[   \sum_k F_k \otimes \mathcal{L}_k - i \omega_L F_z   \right] \vec{\rho} = 0 \,,
\end{equation}
with $\vec{\rho} $ a vector containing all Fourier components of $\rho(t)$.

\section{Time-Independent Interaction} \label{App:Benchmark}

\begin{figure}[t!]
\begin{center}
  \includegraphics[width=.99\linewidth]{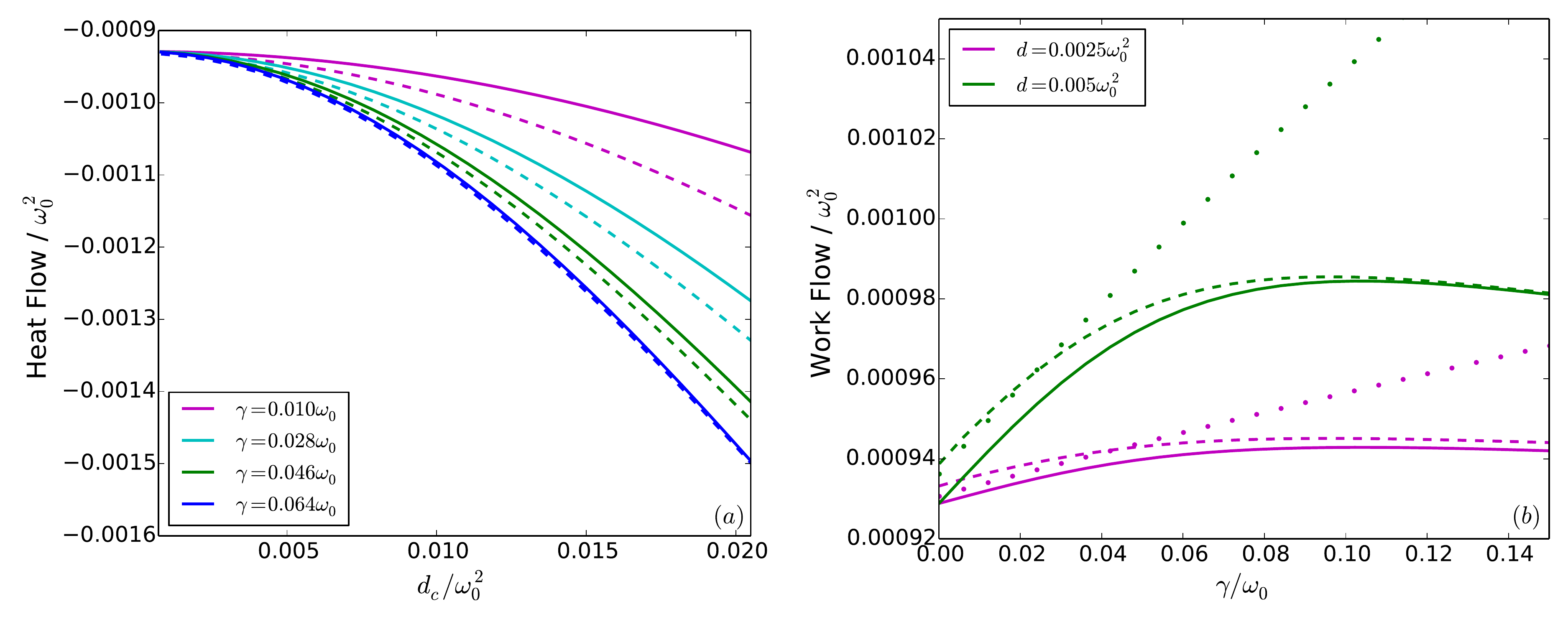}
  \end{center}
\caption{(a) Heat flow from the hot reservoir as a function of the coupling strength. (b) Work flow as a function of parameter $\gamma$.
Continuous lines are results of the Markovian treatment of the two level system and dashed lines results obtained using the CC mapping. Parameters used are: $\omega_L = 0.75 \omega_0, \, g = 0.2 \omega_0,\,  d_h = 0.005\omega_0^2, \, \beta_c\omega_0= 25, \beta_h\omega_0 = 2.2$ and $ \omega_{res} = 0.2 \omega_0$. In (b) dots indicate results using the CC mapping with the secular approximation.}
\label{Fig:BM}
\end{figure}
We consider the case where the working medium is coupled to the cold bath and hot bath with the same system 
coupling operator $S_c = S_h = \sigma_x /\sqrt{ 2 \omega_0}$. Note that we have removed the time dependence in the coupling operator that 
appeared in our original model. This is done in order to compare results of the heat flows calculated with and without the CC mapping. The overall coupling strength between system and reservoir is controlled by the parameter $d_c$ in (\ref{eq:CSD}). Parameter $\gamma$, on the other hand, can tune how strongly peaked the spectral density is. A smaller value of $\gamma$ means a more strongly peaked spectral density. 

Figure~\ref{Fig:BM}~(a) shows the heat flow  of the hot reservoir as a function of the coupling strength for different 
 values of $\gamma$. Continuous lines are the result of a standard Markovian theory, where the Born, Markov and secular 
 approximations have been performed and dashed lines show the result of the non secular master equation of the supersystem obtained after the CC mapping. We can clearly see that the 
 structure of the spectral density can have a strong impact on steady-state quantities, since the smaller 
 the value of $\gamma$ the bigger the difference between the standard Markovian theory and the results performing the 
 CC mapping. The reason for this relies in the separation of timescales typically performed during the Markovian 
 approximation. In this approximation, it is assumed that the correlation functions of the reservoir decay at a much 
 faster rate than the typical timescales of the system. As the spectral density gets more peaked, correlation functions 
 tend to decay at a slower rate, thus making the separation of timescales not valid. The heat flow of the cold reservoir and the work flow also follow a similar behavior as a function of the coupling strength for different values of $\gamma$.

Figure~\ref{Fig:BM}~(b) shows the work flow as a function of $\gamma$ for small couplings $d_c$. This time we have included 
 the results performing the CC mapping with the secular approximation (dots) for completeness. As explained in Sec. 
 \ref{sec:Floquet}, the study of a periodically driven system with Floquet theory can make the secular approximation not 
 suitable for analysis \cite{Hone2009}. The secular approximation requires the system’s level broadening to be much smaller than its level spacing. 
 After the CC mapping is performed the dimension of the supersystem becomes infinite. Since for a correct treatment of 
 the driven problem all quasienergies (see~\ref{App:Floquet theory}) are mapped to the same Brillouin zone, the 
 assumption regarding the system's level spacing looses its validity. Figure~\ref{Fig:BM}~(b) shows that the results of performing 
 the CC mapping with the secular approximation might agree with the results of only performing the CC 
 mapping in some regimes, nevertheless  to fully identify this regimes independently of the model, a further study is needed. As in figure~\ref{Fig:BM}(a) we also see that bigger differences between the standard Markoivan theory and the CC mapping 
formalism are obtained for strongly peaked spectral densities. Similar results to the ones in figure~\ref{Fig:BM}~(a) were also obtained in \cite{Strasberg2016} for the case of a non-driven heat engine.

 We stress the point that even though higher values of $\gamma$ show a better agreement with the standard Markovian theory after the CC mapping is performed, the coupling between supersystem and residual reservoir must remain weak in order to obtain a master equation. Since in our model the mapped spectral density is proportional to $\gamma$ (see~\ref{App:CC}), one needs to be careful because increasing the value of $\gamma$ might make the master equation for the supersystem not valid.

\section{Laser Cooling} \label{App:Laser Cooling}
 We consider an ion interacting with a laser field \cite{Cirac1992,Moya-Cessa2012}. The interaction is given by $H_I=\vec{d}\cdot \vec{E}$, where $\vec{d}$ is the dipole moment and $\vec{E}$ the electric field. We consider only two electronic states of the ion and define the electric field to be along the x-direction,
\begin{equation}
 H_I = d \vec{S} \cdot \vec{E}(\vec{r}, t) = d \vec{S} \cdot \vec{E} \cos(kx - \omega t) \, ,
\end{equation}
with $\vec{S}$ the spin of the ion and x its position. We also assume that the motional degree of freedom of the ion is 
quantized in a harmonic potential such that $x = x_0 (a + a^{\dagger})$ and define $\eta = kx_0$, so that
\begin{equation}
 H_I  = \frac{d E }{2} \sigma_x \cos \left[  \eta (a + a^{\dagger}) - \omega t    \right] \, .
\end{equation}
If the ion is well confined we can assume that $\eta \ll 1$. Defining $\Omega \equiv d E / 2$ we obtain
\begin{equation}
 H_I  = \Omega \,  \sigma_x \cos (\omega t ) - \Omega \,\eta \,  \sigma_x \, (a+a^{\dagger}) \sin(\omega t)  \, ,
\end{equation}
which is analogous to the time dependent part of the mapped working medium in equation~(\ref{eq:Super_H}).

\section{Fluctuations} \label{App:Fluctuations}
Here we analyze the fluctuations in the occupation number of the CC, defined as $ \langle (\Delta n)^2 \rangle \equiv \langle n ^2 \rangle   - \langle n \rangle  ^ 2 $, with $\langle n\rangle = \langle a^{\dagger}a \rangle$. For a thermal state (th) it is easy to show that these fulfil $ \langle (\Delta n)^2 \rangle_{th} = \langle n  \rangle_{th} ( 1  + \langle n \rangle_{th} )\geq \langle n  \rangle_{th} $.
\begin{figure}[ht]
\begin{center}
  \includegraphics[width=0.5\linewidth]{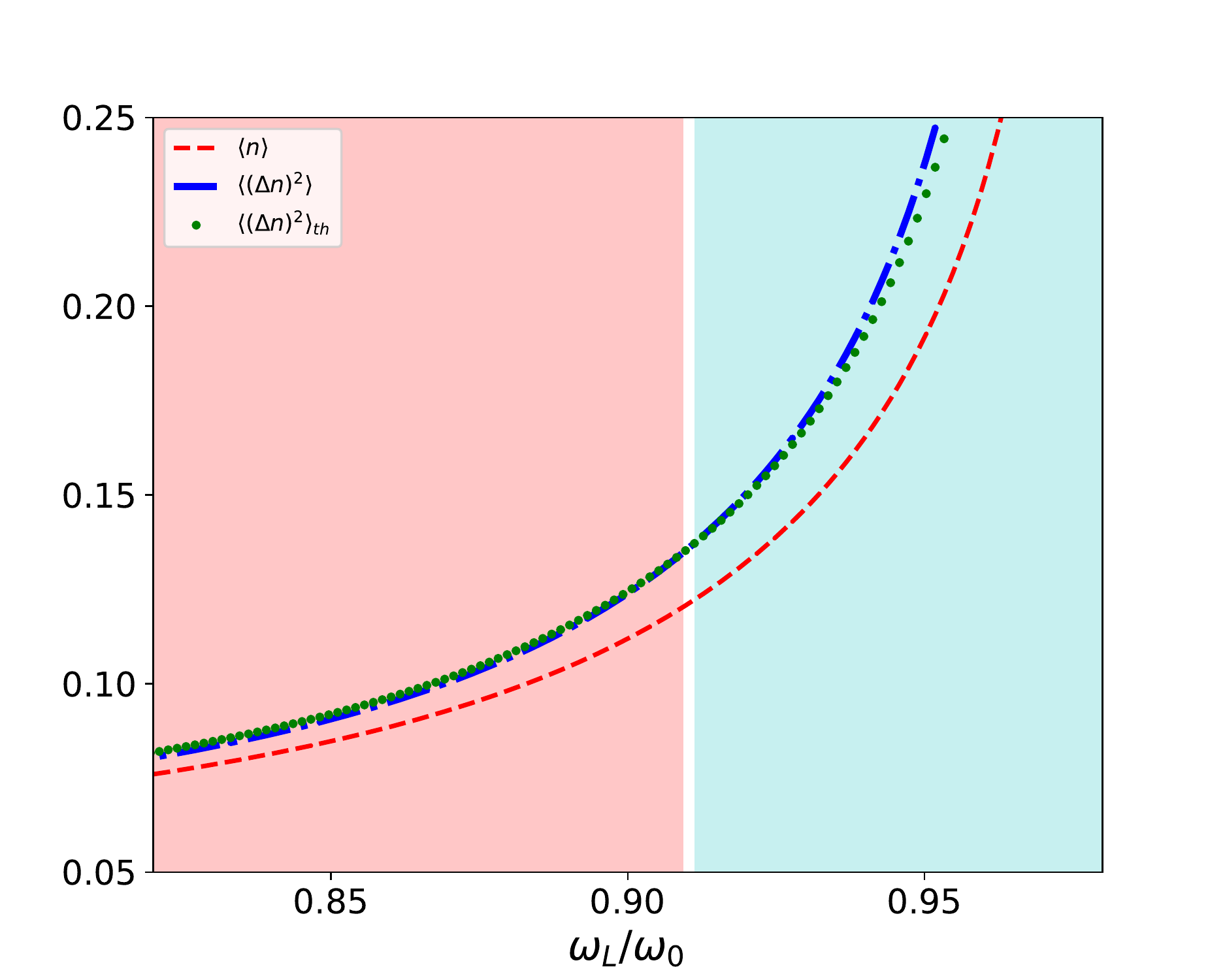}
\end{center}    
\caption{Left: Occupation number of the CC and its fluctuations as a function of the driving frequency $\omega_L$ while keeping the resonance 
condition (20). Parameters used are as in figure 2.}
\label{Fig:n_fluc}
\end{figure}

Figure \ref{Fig:n_fluc} shows a good agreement between the variance of the occupation number of the CC and that of a thermal state, further supporting the choice of parametrisation for the effective temperature in equation (\ref{eq:betaCC}).

\section*{References}
\bibliographystyle{iopart-num}

\bibliography{Mybib_new}

\providecommand{\newblock}{}
\begin{thebibliography}{10}
\expandafter\ifx\csname url\endcsname\relax
  \def\url#1{{\tt #1}}\fi
\expandafter\ifx\csname urlprefix\endcsname\relax\def\urlprefix{URL }\fi
\providecommand{\eprint}[2][]{\url{#2}}

\bibitem{Hugel2002}
Hugel T, Holland N~B, Cattani A, Moroder L, Seitz M and Gaub H~E 2002 {\em
  Science\/} {\bf 296} 1103--6

\bibitem{Pekola2015}
Pekola J~P 2015 {\em Nat. Phys.\/} {\bf 11} 118

\bibitem{Martinez2016}
Mart{\'{i}}nez I~A, Rold{\'{a}}n E, Dinis L, Petrov D, Parrondo J~M and Rica
  R~A 2016 {\em Nat. Phys.\/} {\bf 12} 67--70

\bibitem{Krishnamurthy2016}
Krishnamurthy S, Ghosh S, Chatterji D, Ganapathy R and Sood A~K 2016 {\em Nat.
  Phys.\/} {\bf 12} 1134--1138

\bibitem{Roßnagel2016}
Ro{\ss}nagel J, Dawkins S~T, Tolazzi K~N, Abah O, Lutz E, Schmidt-Kaler F and
  Singer K 2016 {\em Science\/} {\bf 352} 325--9

\bibitem{Maslennikov2017}
Maslennikov G, Ding S, Hablutzel R, Gan J, Roulet A, Nimmrichter S, Dai J,
  Scarani V and Matsukevich D 2017  (\textit{Preprint} \eprint{1702.08672})

\bibitem{Klatzow2017}
Klatzow James;~Weinzetl C, Ledingham P~M, Becker J~N, Saunders D~J, Nunn J,
  Walmsley I~A, Uzdin R and Poem E 2017  (\textit{Preprint}
  \eprint{1710.08716})

\bibitem{Spohn1978a}
Spohn H and Lebowitz J~L 1978 {\em Adv. Chem. Phys.\/} {\bf 38} 109

\bibitem{Alicki1979}
Alicki R 1979 {\em J. Phys. A. Math. Gen.\/} {\bf 12} L103--L107

\bibitem{Kosloff2014}
Kosloff R and Levy A 2014 {\em Annu. Rev. Phys. Chem.\/} {\bf 65} 365--393

\bibitem{Callen1985}
Callen H~B 1985 {\em {Thermodynamics and an Introduction to
  Thermostatistics}\/} (Singapore: John Wiley {\&} Son)

\bibitem{Shirley1965}
Shirley J~H 1965 {\em Phys. Rev.\/} {\bf 138} B979--B987

\bibitem{Sambe1973}
Sambe H 1973 {\em Phys. Rev. A\/} {\bf 7} 2203--2213

\bibitem{Grifoni1998}
Grifoni M and H{\"{a}}nggi P 1998 {\em Phys. Rep.\/} {\bf 304} 229--354

\bibitem{Grossmann1991}
Grossmann F, Dittrich T, Jung P and H{\"{a}}nggi P 1991 {\em Phys. Rev.
  Lett.\/} {\bf 67} 516--519

\bibitem{Grossmann1992}
Grossmann F and H{\"{a}}nggi P 1992 {\em Europhys. Lett.\/} {\bf 18} 571--576

\bibitem{Dunlap1986}
Dunlap D~H and Kenkre V~M 1986 {\em Phys. Rev. B\/} {\bf 34} 3625--3633

\bibitem{Eckardt2005}
Eckardt A, Weiss C and Holthaus M 2005 {\em Phys. Rev. Lett.\/} {\bf 95} 260404

\bibitem{Lindner2011}
Lindner N~H, Refael G and Galitski V 2011 {\em Nat. Phys\/} {\bf 7} 490--495

\bibitem{Bastidas2012}
Bastidas V~M, Emary C, Regler B and Brandes T 2012 {\em Phys. Rev. Lett.\/}
  {\bf 108} 043003

\bibitem{Kohler1997}
Kohler S, Dittrich T and H{\"{a}}nggi P 1997 {\em Phys. Rev. E\/} {\bf 55} 300

\bibitem{Gelbwaser-Klimovsky2013}
Gelbwaser-Klimovsky D, Alicki R and Kurizki G 2013 {\em Phys. Rev. E\/} {\bf
  87}(1) 012140

\bibitem{Kosloff2013}
Kosloff R 2013 {\em Entropy\/} {\bf 15} 2100--2128

\bibitem{Szczygielski2013}
Szczygielski K, Gelbwaser-Klimovsky D and Alicki R 2013 {\em Phys. Rev. E\/}
  {\bf 87}(1) 012120

\bibitem{Szczygielski2014}
Szczygielski K 2014 {\em J. Math. Phys.\/} {\bf 55} 083506

\bibitem{Gelbwaser-Klimovsky2015a}
Gelbwaser-Klimovsky D, Szczygielski K, Vogl U, Sa\ss{} A, Alicki R, Kurizki G
  and Weitz M 2015 {\em Phys. Rev. A\/} {\bf 91} 023431

\bibitem{Breuer2002}
Breuer H~P and Petruccione F 2002 {\em {The Theory of Open Quantum Systems}\/}
  (Oxford: Oxford University Press)

\bibitem{Shirai2015}
Shirai T, Mori T and Miyashita S 2015 {\em Phys. Rev. E\/} {\bf 91} 030101(R)

\bibitem{Cuetara2015}
Bulnes~Cuetara G, Engel A and Esposito M 2015 {\em New J. Phys.\/} {\bf 17}
  055002

\bibitem{Vogl2009}
Vogl U and Weitz M 2009 {\em Nature\/} {\bf 461} 70--73

\bibitem{Vogl2011}
Vogl U, Sa A, Haelmann S and Weitz M 2011 {\em J. Mod. Opt.\/} {\bf 58}
  1300--1309

\bibitem{Schmidt2015}
Schmidt R, Carusela M~F, Pekola J~P, Suomela S and Ankerhold J 2015 {\em Phys.
  Rev. B\/} {\bf 91} 224303

\bibitem{Restrepo2016}
Restrepo S, Cerrillo J, Bastidas V~M, Angelakis D~G and Brandes T 2016 {\em
  Phys. Rev. Lett.\/} {\bf 117}(25) 250401

\bibitem{Carrega2016}
Carrega M, Solinas P, Sassetti M and Weiss U 2016 {\em Phys. Rev. Lett.\/} {\bf
  116} 240403

\bibitem{Esposito2015}
Esposito M, Ochoa M~A and Galperin M 2015 {\em Phys. Rev. Lett.\/} {\bf 114}
  080602

\bibitem{Bruch2016}
Bruch A, Thomas M, Kusminskiy S~V, von Oppen F and Nitzan A 2016 {\em Phys.
  Rev. B\/} {\bf 93} 115318

\bibitem{Strasberg2016}
Strasberg P, Schaller G, Lambert N and Brandes T 2016 {\em New J. Phys.\/} {\bf
  18} 073007

\bibitem{Newman2017}
Newman D, Mintert F and Nazir A 2017 {\em Phys. Rev. E\/} {\bf 95} 032139

\bibitem{Strasberg2017a}
Strasberg P and Esposito M 2017 {\em Phys. Rev. E\/} {\bf 95} 062101

\bibitem{Schaller2017}
Schaller G, Cerrillo J, Engelhardt G and Strasberg P 2018 {\em Phys. Rev. B\/}
  {\bf 97}(19) 195104

\bibitem{Strasberg2017}
Strasberg P, Schaller G, Schmidt T~L and Esposito M 2018 {\em Phys. Rev. B\/}
  {\bf 97}(20) 205405

\bibitem{Mu2017}
Mu A, Agarwalla B~K, Schaller G and Segal D 2017 {\em New J. Phys.\/} {\bf 19}
  123034

\bibitem{Gallego2014}
Gallego R, Riera A and Eisert J 2014 {\em New J. Phys.\/} {\bf 16} 125009

\bibitem{Perarnau-Llobet2017}
Perarnau-Llobet M, Wilming H, Riera A, Gallego R and Eisert J 2018 {\em Phys.
  Rev. Lett.\/} {\bf 120}(12) 120602

\bibitem{Esposito2010}
Esposito M, Lindenberg K and den Broeck C~V 2010 {\em New J. Phys.\/} {\bf 12}
  013013

\bibitem{Gelbwaser-Klimovsky2015}
Gelbwaser-Klimovsky D and Aspuru-Guzik A 2015 {\em J. Phys. Chem. Lett.\/} {\bf
  6} 3477--3482

\bibitem{Kato2016}
Kato A and Tanimura Y 2016 {\em J. Chem. Phys.\/} {\bf 145} 224105

\bibitem{Cerrillo2016}
Cerrillo J, Buser M and Brandes T 2016 {\em Phys. Rev. B\/} {\bf 94} 214308

\bibitem{Martinazzo2011}
Martinazzo R, Vacchini B, Hughes K~H and Burghardt I 2011 {\em J. Chem.
  Phys.\/} {\bf 134}

\bibitem{Woods2014}
Woods M~P, Groux R, Chin A~W, Huelga S~F and Plenio M~B 2014 {\em J. Math.
  Phys.\/} {\bf 55} 032101

\bibitem{Esposito2009}
Esposito M, Harbola U and Mukamel S 2009 {\em Rev. Mod. Phys.\/} {\bf 81}
  1665--1702

\bibitem{Garg1985}
Garg A, Onuchic J~N and Ambegaokar V 1985 {\em J. Chem. Phys.\/} {\bf 83}
  4491--4503

\bibitem{Iles-Smith2014}
Iles-Smith J, Lambert N and Nazir A 2014 {\em Phys. Rev. A\/} {\bf 90} 032114

\bibitem{Iles-Smith2016}
Iles-Smith J, Dijkstra A~G, Lambert N and Nazir A 2016 {\em J. Chem. Phys.\/}
  {\bf 144} 044110

\bibitem{Martinazzo2011a}
Martinazzo R, Hughes K~H and Burghardt I 2011 {\em Phys. Rev. E\/} {\bf 84}(3)
  030102

\bibitem{Prior2010}
Prior J, Chin A~W, Huelga S~F and Plenio M~B 2010 {\em Phys. Rev. Lett.\/} {\bf
  105}(5) 050404

\bibitem{Chin2010}
Chin A~W, Rivas A, Huelga S~F and Plenio M~B 2010 {\em J. Math. Phys.\/} {\bf
  51} 092109

\bibitem{Rosenbach2016}
Rosenbach R, Cerrillo J, Huelga S~F, Cao J and Plenio M~B 2016 {\em New J.
  Phys.\/} {\bf 18} 023035

\bibitem{huh2014a}
Huh J, Mostame S, Fujita T, Yung M~H and Aspuru-Guzik A 2014 {\em New J.
  Phys.\/} {\bf 16} 123008

\bibitem{Hone2009}
Hone D~W, Ketzmerick R and Kohn W 2009 {\em Phys. Rev. E\/} {\bf 79} 051129

\bibitem{Cirac1992}
Cirac J~I, Blatt R, Zoller P and Phillips W~D 1992 {\em Phys. Rev. A\/} {\bf
  46} 2668--2681

\bibitem{Eckardt2015}
Eckardt A and Anisimovas E 2015 {\em New J. Phys.\/} {\bf 17} 093039

\bibitem{Moya-Cessa2012}
Moya-Cessa H, Soto-Eguibar F, Vargas-Martínez J~M, Juárez-Amaro R and
  Zúñiga-Segundo A 2012 {\em Phys. Rep.\/} {\bf 513} 229 -- 261

\end{thebibliography}

\end{document}